\documentclass[twocolumn]{aastex631}

\shorttitle{Eclipse Mapping of WASP-43b}
\shortauthors{Hammond et al.}

\graphicspath{{./}}

\begin{document}

\title{Two-Dimensional Eclipse Mapping of the Hot Jupiter WASP-43b with JWST MIRI/LRS}

\author[0000-0002-6893-522X]{Mark Hammond}
\affiliation{Atmospheric, Oceanic, and Planetary Physics, Department of Physics, University of Oxford, Oxford, UK}
\author[0000-0003-4177-2149]{Taylor J.\ Bell}
\affiliation{BAER Institute, NASA Ames Research Center, Moffet Field, CA, USA}
\affiliation{Space Science and Astrobiology Division, NASA Ames Research Center, Moffett Field, CA, USA}
\author[0000-0002-8211-6538]{Ryan C.\ Challener}
\affiliation{Department of Astronomy, Cornell University, Ithaca, NY, USA}
\affiliation{Department of Astronomy, University of Michigan, Ann Arbor, MI, USA}
\author{Neil T.\ Lewis}
\affiliation{Department of Mathematics and Statistics, University of Exeter, Exeter, Devon, UK}
\author{Megan Weiner Mansfield}
\affiliation{Steward Observatory, University of Arizona, Tucson, AZ, USA}
\affiliation{NHFP Sagan Fellow}
\author{Isaac Malsky}
\affiliation{Department of Astronomy, University of Michigan, Ann Arbor, MI, USA}
\author[0000-0003-3963-9672]{Emily Rauscher}
\affiliation{Department of Astronomy, University of Michigan, Ann Arbor, MI, USA}
\author[0000-0003-4733-6532]{Jacob L.\ Bean}
\affiliation{Department of Astronomy \& Astrophysics, University of Chicago, Chicago, IL, USA}
\author{Ludmila Carone}
\affiliation{Space Research Institute, Austrian Academy of Sciences, Graz, Austria}
\author[0000-0002-6907-4476]{Jo\~ao M.\ Mendon\c ca}
\affiliation{DTU Space, Technical University of Denmark, Kgs. Lyngby, Denmark}
\author[0000-0002-0797-5746]{Lucas Teinturier}
\affiliation{LESIA, Observatoire de Paris, Universit\'{e} PSL, Sorbonne Université, Universit\'{e} Paris Cit\'{e}, CNRS, Meudon, France}
\affiliation{Laboratoire de Météorologie Dynamique, IPSL, CNRS, Sorbonne Université, Ecole Normale Supérieure, Université PSL, Ecole Polytechnique, Institut Polytechnique de Paris, Paris, France
}
\author{Xianyu Tan}
\affiliation{Tsung-Dao Lee Institute, Shanghai Jiao Tong University, 520 Shengrong Road, Shanghai, People’s Republic of China}
\affiliation{School of Physics and Astronomy, Shanghai Jiao Tong University, 800 Dongchuan Road, Shanghai, People’s Republic of China}

\author[0000-0001-7866-8738]{Nicolas Crouzet}
\affiliation{Leiden Observatory, Leiden University, Leiden, The Netherlands}
\author{Laura Kreidberg}
\affiliation{Max Planck Institute for Astronomy, Heidelberg, Germany}
\author[0000-0002-4262-5661]{Giuseppe Morello}
\affiliation{Department of Space, Earth and Environment, Chalmers University of Technology, Gothenburg, Sweden}
\affiliation{Instituto de Astrof\'isica de Canarias (IAC), La Laguna, Tenerife, Spain}
\affiliation{Instituto de Astrof\'isica de Andaluc\'ia (IAA-CSIC), Glorieta de la Astronom\'ia s/n, 18008 Granada, Spain}
\author{Vivien Parmentier}
\affiliation{Université Côte d’Azur, Observatoire de la Côte d’Azur, CNRS, Laboratoire Lagrange, France}

\author[0000-0002-0769-9614]{Jasmina Blecic}
\affiliation{Department of Physics, New York University Abu Dhabi, Abu Dhabi, UAE}
\affiliation{Center for Astro, Particle and Planetary Physics (CAP3), New York University Abu Dhabi, Abu Dhabi, UAE}
\author{Jean-Michel D\'esert}
\affiliation{Anton Pannekoek Institute for Astronomy, University of Amsterdam, Science Park 904, 1098 XH, Amsterdam, The Netherlands}
\author{Christiane Helling}
\affiliation{Space Research Institute, Austrian Academy of Science, Schmiedlstrasse 6, 8042 Graz, Austria}
\affiliation{Institute for Theoretical Physics and Computational Physics, Graz University of Technology, Petersgasse16/II, 8010 Graz, Austria}
\author{Pierre-Olivier Lagage}
\affiliation{Université Paris-Saclay, Université Paris Cité, CEA, CNRS, AIM, Gif-sur-Yvette, France}
\author{Karan Molaverdikhani}
\affiliation{Universitäts-Sternwarte, Ludwig-Maximilians-Universität München, Scheinerstrasse 1, D-81679 München, Germany}
\affiliation{Exzellenzcluster Origins, Boltzmannstraße 2, 85748 Garching, Germany}
\author[0000-0001-8236-5553]{Matthew C. Nixon}
\affiliation{Department of Astronomy, University of Maryland, College Park, Maryland, USA}
\author{Benjamin V.\ Rackham}
\affiliation{Department of Earth, Atmospheric and Planetary Sciences, Massachusetts Institute of Technology, Cambridge, MA, USA}
\affiliation{Kavli Institute for Astrophysics and Space Research, Massachusetts Institute of Technology, Cambridge, MA, USA}
\author[0009-0006-2395-6197]{Jingxuan Yang}
\affiliation{Atmospheric, Oceanic, and Planetary Physics, Department of Physics, University of Oxford, Oxford, UK}

\begin{abstract}
We present eclipse maps of the two-dimensional thermal emission from the dayside of the hot Jupiter WASP-43b, derived from an observation of a phase curve with the JWST MIRI/LRS instrument. The observed eclipse shapes deviate significantly from those expected for a planet emitting uniformly over its surface. We fit a map to this deviation, constructed from spherical harmonics up to order $\ell_{\rm max}=2$, alongside the planetary, orbital, stellar, and systematic parameters. This yields a map with a meridionally-averaged eastward hot-spot shift of $(7.75 \pm 0.36)^{\circ}$, with no significant degeneracy between the map and the additional parameters. We show the latitudinal and longitudinal contributions of the day-side emission structure to the eclipse shape, finding a latitudinal signal of {$\sim$}200 ppm and a longitudinal signal of {$\sim$}250 ppm. To investigate the sensitivity of the map to the method, we fix the non-mapping parameters and derive an ``eigenmap'' fitted with an optimised number of orthogonal phase curves, which yields a similar map to the $\ell_{\rm max}=2$ map. We also fit a map up to $\ell_{\rm max}=3$, which shows a smaller hot-spot shift, with a larger uncertainty. These maps are similar to those produced by atmospheric simulations. We conclude that there is a significant mapping signal which constrains the spherical harmonic components of our model up to $\ell_{\rm max}=2$. Alternative mapping models may derive different structures with smaller-scale features; we suggest that further observations of WASP-43b and other planets will drive the development of more robust methods and more accurate maps.
\end{abstract}


\section{Introduction} \label{sec:intro}

Eclipse mapping measures the two-dimensional spatial features of an object when it is eclipsed by another object. The eclipsed object is covered along one axis during the eclipse ingress, and is revealed along another axis during the eclipse egress. Combining the information from these two axes reveals two-dimensional information about the surface of the eclipsed object \citep{de2012towards,majeau2012two}.

Eclipse maps have been derived for objects like Pluto during its eclipse by Charon \citep{stern1992pluto}, or the white dwarf BD +16$^{\circ}$516B during its eclipse in a binary system \citep{warner1971measurement}. \citet{williams2006resolving} proposed applying this technique to exoplanets, and \citet{rauscher2007toward} showed that the James Webb Space Telescope (JWST) would have sufficient precision to accurately map exoplanets. Eclipse mapping is currently the only method by which 2D information can be measured for exoplanets, as out-of-eclipse ``phase curves'' provide low-resolution information as a function of longitude only. 2D spatial information is crucial for understanding atmospheric circulation \citep{lewis2022temperature}, chemical composition \citep{taylor2020understanding,yang2023testing}, and cloud structure \citep{parmentier2016transitions}.

\citet{de2012towards} and \citet{majeau2012two} derived eclipse maps of the hot Jupiter HD 189733b using observations with the \textit{Spitzer Space Telescope}. Their analyses were restricted to fitting large-scale shapes with a high degree of uncertainty due to the limited precision of the measurements, uncertainty in and degeneracy with orbital parameters, and the dependence of the result on the mapping model. \citet{de2012towards} described the uncertainty in the mapping signal due to the impact parameter, stellar density, eccentricity, and argument of periastron in particular. \citet{rauscher2018eigenmap}, \citet{mansfield2020eigenspectra}, and \citet{challener2022theresa} developed more advanced methods to fit eclipse maps of 2D and 3D structure, based on fitting orthogonal phase curves rather than orthogonal surface maps, and applied these methods to the previous observations of HD 189733b. \citet{coulombe2023wasp18b} presented the first eclipse map measured with JWST, deriving a map of thermal emission from observations of the hot Jupiter WASP-18b from 0.85 to 2.85 $\mu$m with the NIRISS/SOSS instrument. They found no longitudinal shift in the hot-spot, sharp gradients in brightness towards the terminator, and no clear measurement of latitudinal structure.

This study presents eclipse maps derived from broadband JWST MIRI/LRS \citep{kendrew2015mid} observations from 5 to 10.5 $\mu$m of a full phase curve of the hot Jupiter WASP-43b containing two eclipses and one transit \citep{bell2023wasp43b}. WASP-43b is a ``hot Jupiter'' exoplanet with strong thermal emission from its permanent dayside, and a short, tidally-locked orbit around a K7 main sequence star that exhibits low variability \citep{hellier2011wasp,scandariato2022phase}. These properties make it ideal for precise time-series observations of thermal emission. 

\citet{bell2023wasp43b} analysed this MIRI/LRS phase curve, finding a large difference in dayside and nightside brightness temperatures and evidence for water absorption. We fit its inclination to be $82.11 ^{+ 0.050 } _{-0.052}$, and the ratio of its semi-major axis to stellar radius to be $4.859^{+0.013}_{-0.012}$, which corresponds to an impact parameter of $0.667^{+0.006}_{-0.006}$. This means that the edge of the star crosses the planet at an angle of $\sim$42$^{\circ}$ (the ``stellar edge angle'' defined in \cite{boone2023analytic}). The longitudinal and latitudinal features of the day-side therefore affect the eclipse mapping signal almost equally, with a small bias towards longitudinal features. This geometry makes WASP-43b especially well suited for eclipse mapping.

We fit an eclipse map to this dataset, finding the clearest eclipse mapping signal to date and separating the latitudinal and longitudinal parts of this signal for the first time. In Section \ref{sec:methods} we describe the methods used to fit the eclipse maps. Section \ref{sec:results} then shows the different maps we fit to the observations, and compares their structures and statistical evidence. These maps are then compared to numerical simulations in Section \ref{sec:discussion}. In Section \ref{sec:conclusions} we conclude that the spherical harmonic components up to $\ell_{\rm max}=2$ are constrained well, but that further observations are needed to precisely constrain higher-order mapping structures.

\begin{figure*}
    \centering
    \includegraphics[width=\textwidth]{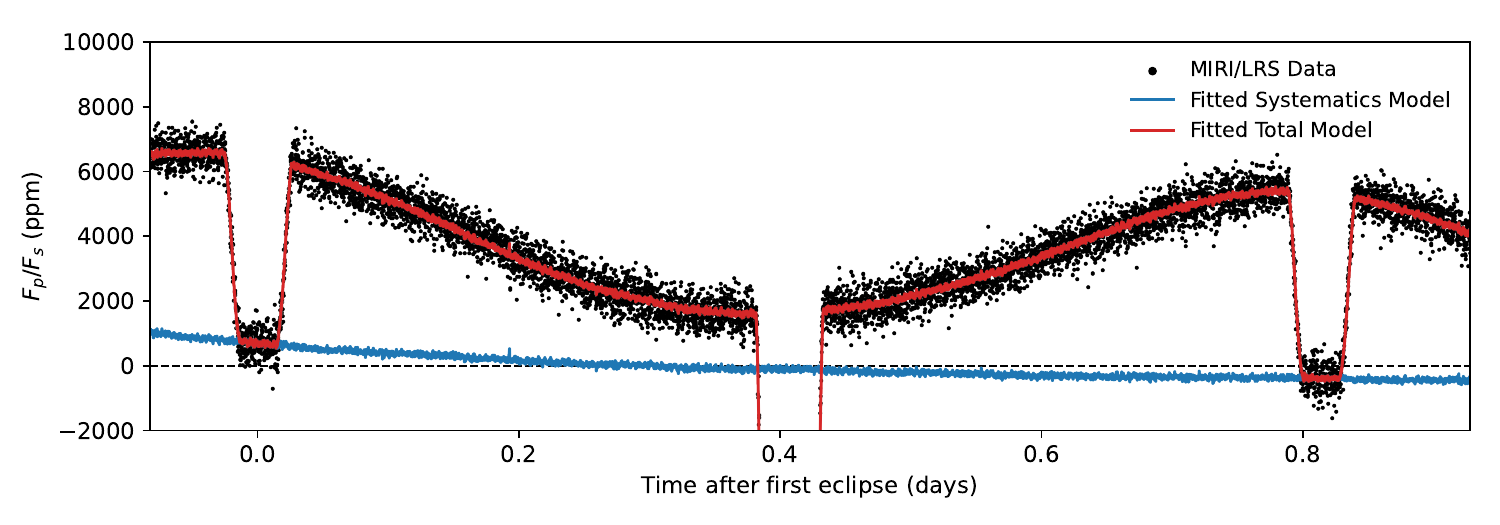}
    \caption{The JWST MIRI/LRS dataset, showing the observed phase curve (black points), the fitted $\ell_{\rm max}=2$ eclipse map model (red line) shown in Figure \ref{fig:fig3_o2_map}, and the systematics model (blue line) fitted alongside the eclipse map.}
    \label{fig:fig1_lightcurve}
% \end{figure*}

% \begin{figure*}
    \centering
    \includegraphics[width=\textwidth]{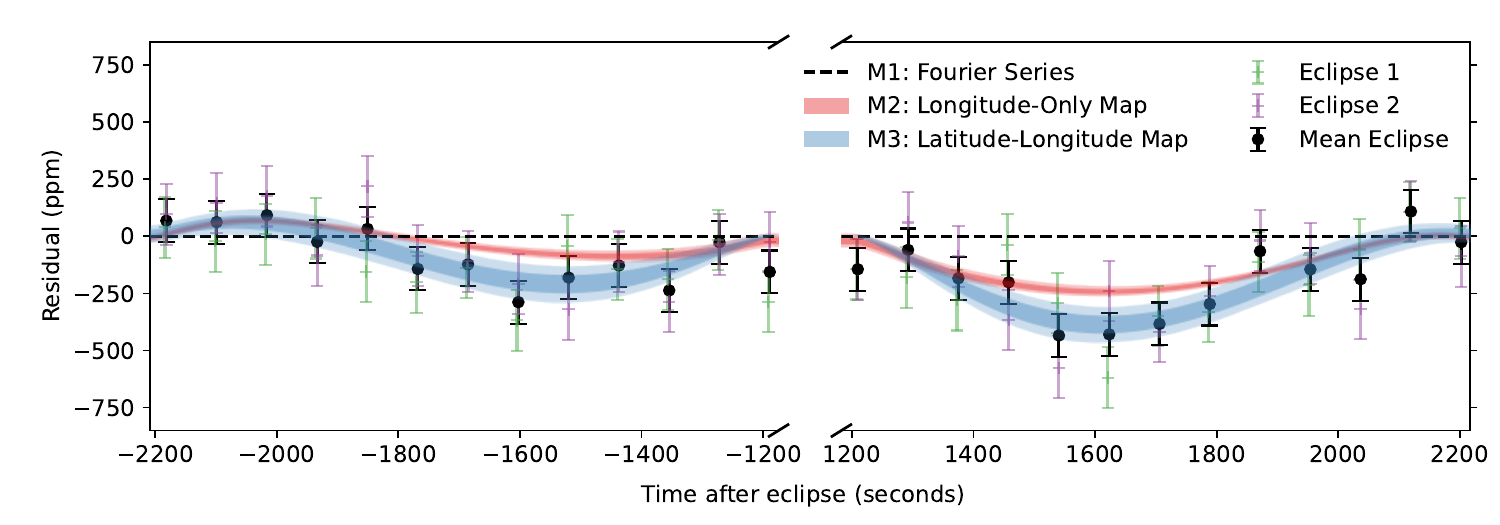}
    \caption{The observed data in the ingress and egress of the eclipses, subtracted by a phase curve fitted using an $n=2$ Fourier series model. We fitted this Fourier series model using the orbital parameters derived using an eclipse map in Table \ref{tab:params_table}, so that the residual signal depends entirely on the different models. This leaves a residual ``eclipse mapping signal'' showing the effect of partial stellar coverage of the non-uniform planetary emission. Green points show the first eclipse, purple points show the second eclipse, and black points show their mean. The blue shaded region labelled ``Latitude-Longitude Map Fit'' shows the range of fitted phase curves from the $\ell_{\rm max}=2$ eclipse map in Figure \ref{fig:fig3_o2_map}, with the two shaded regions showing the first and second quantiles, containing $68.27\%$ and $95.45\%$ of the posterior distribution. The red shaded region labelled ``Longitude-Only Map Fit'' shows the fitted eclipse of the $\ell_{\rm max}=2$ eclipse map with flat latitudinal structure shown in Figure \ref{fig:fig4_o2_flat_map}. The poor fit of this model shows the presence of latitudinal information in the dataset, and the need to fit the latitudinal structure of the map. This ``longitude-only'' residual is around 250 ppm at its largest, while the ``latitude-only'' signal (estimated from the difference between the 1D and 2D map fits) is around 200 ppm. The total  eclipse mapping signal is around 450 ppm at its largest.}
    \label{fig:fig2_residual}
\end{figure*}

\begin{table*}
    \centering
    \begin{tabular}{c|cc}
        \textbf{Parameter} & \textbf{Fourier Series Fit} & \textbf{Eclipse Map Fit}\\
        \hline
Planet-Star Radius Ratio $R_{p}$ ($R_{*}$) & $ 0.15734 ^{+ 0.00017 } _{- 0.00015 }$ & $ 0.15839 ^{+ 0.00025 } _{- 0.00040 }$ \\
Transit Time (BMJD) & $ 55934.292296 ^{+ 1.2 \times 10^{-5} } _{- 1.2 \times 10^{-5} }$ & $ 55934.292283 ^{+ 1.1 \times 10^{-5} } _{- 1.1 \times 10^{-5} }$ \\
Inclination ($^{\circ}$) & $ 82.277 ^{+ 0.050 } _{- 0.050 }$ & $ 82.106 ^{+ 0.050 } _{- 0.052 }$ \\
Semi-Major Axis $a$ ($R_{*}$) & $ 4.881 ^{+ 0.012 } _{- 0.012 }$ & $ 4.859 ^{+ 0.013 } _{- 0.012 }$ \\
Limb Darkening Parameter $q_1$ & $ 0.0565 ^{+ 0.0079 } _{- 0.0077 }$ & $ 0.0182 ^{+ 0.0081 } _{- 0.0045 }$ \\
Limb Darkening Parameter $q_2$ & $ 0.043 ^{+ 0.068 } _{- 0.032 }$ & $ 0.595 ^{+ 0.280 } _{- 0.350 }$ \\
Constant Baseline $C_0$ (ppm)  & $ -2577.0 ^{+ 37 } _{- 50 }$ & $ -2881.0 ^{+ 30 } _{- 30 }$ \\
Linear Trend $C_1$ (ppm/day) & $ -910 ^{+ 100 } _{- 80 }$ & $ -240 ^{+ 60 } _{- 60 }$ \\
Ramp Magnitude $r_0$ (ppm) & $ 758.0 ^{+ 80 } _{- 63 }$ & $ 1319.0 ^{+ 65 } _{- 67 }$ \\
Ramp Time Constant $r_1$ (1/day) & $ 9.7 ^{+ 2.0 } _{- 1.9 }$ & $ 3.7 ^{+ 0.3 } _{- 0.3 }$ \\
Spatial Position Trend & $ 0.0121 ^{+ 0.0012 } _{- 0.0012 }$ & $ 0.0122 ^{+ 0.0012 } _{- 0.0013 }$ \\
Spatial PSF Width Trend & $ -0.0362 ^{+ 0.0070 } _{- 0.0069 }$ & $ -0.0385 ^{+ 0.0071 } _{- 0.0072 }$ \\
Uncertainty Scaling Factor & $ 1.2252 ^{+ 0.0098 } _{- 0.0093 }$ & $ 1.2225 ^{+ 0.0092 } _{- 0.0092 }$ \\
    \end{tabular}
    \caption{The orbital and systematic parameters shown in Figure \ref{fig:fig10_corner}, for the Fourier series fit described in Section \ref{sec:results} and the $\ell_{\rm max}=2$ eclipse map fit plotted in Figure \ref{fig:fig3_o2_map}. The limb darkening parameters $q_1$ and $q_2$ are as described in \citet{kipping2013efficient}. The polynomial parameters $C_0$ and $C_1$ describe the constant baseline and linear trend with time, as used in \citet{bell2022eureka}. The ramp parameters describe the magnitude and timescale of a linearly decaying exponential $r_{0} e^{(-r_{1}t )}$ (where $t=0$ at the start of the observation) as used in \citet{bell2022eureka}. The spatial position and PSF width are as described in \citet{bell2022eureka}. The uncertainty scaling describes a multiplicative parameter to inflate the expected errors to be consistent with the residual between the data and the fitted model.}
    \label{tab:params_table}
\end{table*}

\section{Data and Methods} \label{sec:methods}

A full orbit of WASP-43b with the JWST MIRI/LRS slitless (SL) mode was observed as part of the JWST-ERS-1366 program, performing target acquisition with the F1500W filter and using the SLITLESSPRISM subarray for the science observation. The science observation lasted 26.5 hours, consisting of 9316 integrations lasting 10.34 s each. The two eclipses show a downwards systematic trend with time, similar to that identified in other MIRI/LRS time-series observations \citep{bouwman2023spectroscopic}. We discard the data beyond 10.5 $\mu$m due to the ``shadowed region effect'' described in \citet{bell2023wasp43b} alongside a more detailed analysis of the data acquisition.

Figure \ref{fig:fig1_lightcurve} shows the raw MIRI/LRS dataset (black points) in units of planetary flux divided by stellar flux. We use the \texttt{Eureka!} pipeline \citep{bell2022eureka} to reduce the observed data, starting from the \texttt{Eureka!} Stage 4 output from the ``Eureka v1'' reduction in \citet{bell2023wasp43b}. We trim the initial 780 integrations (not plotted) where instrumental systematic effects are the strongest. The only subsequent methodological difference to \citet{bell2023wasp43b} is the method used to model the planetary emission in the phase curve --- we used a \texttt{starry} model of a planet with a 2D emission map over its surface \citep{luger2019starry}, instead of a Fourier series model of the phase curve.

\subsection{Astrophysical Model}\label{sec:methods:astro_model}

We fit a 2D map of thermal emission to this phase curve, simultaneously with the orbital, planetary, stellar, and instrumental systematic parameters. Given the significant effect of the planetary emission map on the observed eclipse shape, this should derive more accurate parameters than those derived using a Fourier series model in \citet{bell2023wasp43b}. This simultaneous fit also tests if there are any degeneracies between the system parameters and the eclipse map, such as between the eclipse timing and the longitudinal emission offset \citep{williams2006resolving}. 

We fit the planet-to-star radius ratio, the linear ephemeris, the inclination, the ratio of the semi-major axis to the stellar radius, and the two parameterised stellar limb-darkening parameters in \citet{kipping2013efficient}.

We set the orbital period constant at 0.813474 days as it is known with sufficient precision already \citep{kokori2023exoclock}. We set the obliquity of the planet to be 0 as we assume a tidally locked orbit. We set the stellar radius to be constant at 0.665\,R$_{\odot}$ \citep{bell2023wasp43b}, although this is an arbitrary value that serves only to give our model a dimensional form, as the ratio $F_{p}/F_{S}$ is only sensitive to the ratios $R_{p}/R_{*}$ and $a/R_{*}$ which we fit separately. The stellar mass does affect the signal \citep{de2012towards} but is entirely determined in our model by the fitted value of $a/R_{*}$ and the fixed value of the orbital period. 

\citet{de2012towards} showed how orbital eccentricity can be degenerate with an eclipse mapping signal. \citet{gillon2012trappist} constrained the eccentricity of WASP-43b to be $0.0035_{-0.0025}^{+0.0060}$, which has generally been used to assume zero eccentricity for this planet \citep{bell2023wasp43b}. To check this assumption, we separately fitted the model including the eccentricity and periastron, finding an eccentricity of $0.001078_{-0.00025}^{0.00185}$, and no meaningful constraint on periastron, as discussed in Appendix \ref{ap:params}. We take this to verify the assumption of zero eccentricity, and proceed to fit the models with eccentricity set to zero. This is consistent with the circularization timescale of 3 Myr that we estimate using \citet{adams2006long} (assuming $Q_{P}=10^{6}$), which is much shorter than the age of the system estimated by \citet{hellier2011wasp} to be $400^{+200}_{-100}$ Myr.

For the instrumental systematic effects, we fit a uniform baseline and linear trend in time, the magnitude and timescale of an exponential ramp, and trends of the spatial position and width of the data on the detector. Table \ref{tab:params_table} shows the new parameters fitted for the $\ell_{\rm max}=2$ eclipse map model, compared to the parameters fitted with an $n=2$ Fourier series model (like in \citet{bell2023wasp43b}).

The difference in the time it takes light to travel from the planet and the star has an important effect on this dataset, so we include it in our model. In eclipse, the light from the planet takes about 8.5 seconds longer to reach the observer than the light from the star. For the average gradient in eclipse ingress and egress of roughly 8 ppm per second (see Figure \ref{fig:fig1_lightcurve}), this corresponds to a maximum effect of around 70 ppm on the eclipse shape. Figure \ref{fig:fig2_residual} shows that this corresponds to about $15\%$ of the maximum deviation in eclipse shape due to the eclipse mapping signal itself.

\subsection{Eclipse Mapping}

The eclipse mapping method works by constructing a map from a number of ``basis maps'', which each have an associated basis phase curve. We match the observed light curve with a sum $F(t)$ constructed from the the basis phase curves $f_{i}(t)$ weighted by coefficients $c_{i}$:

\begin{equation}\label{eqn:fit-curve}
F(t) = \sum_{i} c_{i} f_{i}(t).
\end{equation}

We derive the actual eclipse map $Z(\theta,\phi)$ from the fitted coefficients $c_{i}$, as a sum of the spherical harmonic basis maps $z_{i}(\theta,\phi)$: 

\begin{equation}
Z(\theta,\phi) = \sum_{i} c_{i} z_{i}(\theta,\phi).
\end{equation}

\citet{luger2019starry} shows the phase curves $f_{i}(t)$ of each spherical harmonic $z_{i}(\theta,\phi)$ (where $\theta$ and $\phi$ are longitude and latitude). We model the astrophysical signal using \texttt{starry}\footnote{\url{starry.readthedocs.io}} \citep{luger2019starry} and fit the coefficients $c_{i}$ using \textit{PyMC3} \citep{salvatier2016probabilistic}. We use ``pixel sampling'' to enforce a positive emission map globally (e.g. as used in \citet{gorski2005healpix}). This method samples the brightness of pixels distributed uniformly over the mapped surface and transforms them to spherical harmonic coefficients $c_{i}$ to compute the actual phase curve. To fit the $\ell_{\rm max}=2$ eclipse map alongside the orbital and systematic parameters, we sample 16 pixels evenly spaced on a Mollweide projection to represent the spherical harmonic space with $4\ell^{2}$ pixels \citep{mcewen2011novel}. We use a (natural) log-normal prior for the pixels with a mean magnitude of 6000 ppm and a standard deviation of 3000 ppm (transformed to log-space).

After fitting the pixels representing the $\ell_{\rm max}=2$ eclipse map simultaneously with the orbital, planetary, stellar, and systematic parameters, we fix these additional parameters to their derived values, and re-fit the map with a variety of methods. We fit maps with:

\begin{enumerate}
    \item $\ell_{\rm max}=3$ spherical harmonics
    \item An eigenmap as described in \citet{rauscher2018eigenmap}
    \item $\ell_{\rm max}=2$ spherical harmonics to the first eclipse alone, the second eclipse alone, and both eclipses
\end{enumerate}

The next section presents the results from each of these re-fitted maps, and describes how the details of each fitting procedure affect the derived map. It also presents the statistical evidence for an eclipse mapping signal, and compares the different mapping models.

\begin{figure*}
    \centering
    \includegraphics[width=\textwidth]{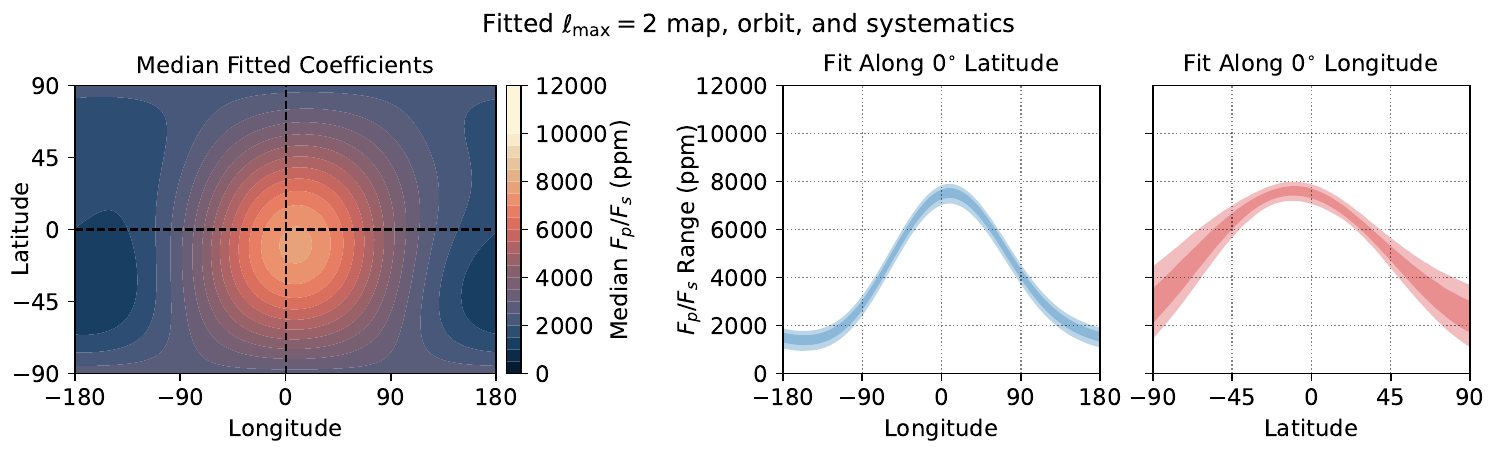}
    \caption{An $\ell_{\rm max}=2$ eclipse map fitted to the data in Figure \ref{fig:fig1_lightcurve}. First panel: an eclipse map constructed with spherical harmonics up to $\ell_{\rm max}=2$, fitted simultaneously with the orbital, planetary, stellar, and systematic parameters, via ``pixel sampling'' as described in Section \ref{sec:results:map}. Note that the nightside of the map contains no latitudinal information, with longitudinal information from the phase curve only (out of eclipse). The plotted 2D map uses the median of the posterior distribution of each fitted spherical harmonic coefficient. Second panel: the posterior distribution of the longitudinal structure along the equator of the map, showing the first and second quantiles. Third panel: the posterior distribution of the latitudinal structure through the substellar point.}
    \label{fig:fig3_o2_map}
% \end{figure*}
% \begin{figure*}
    \centering
    \includegraphics[width=\textwidth]{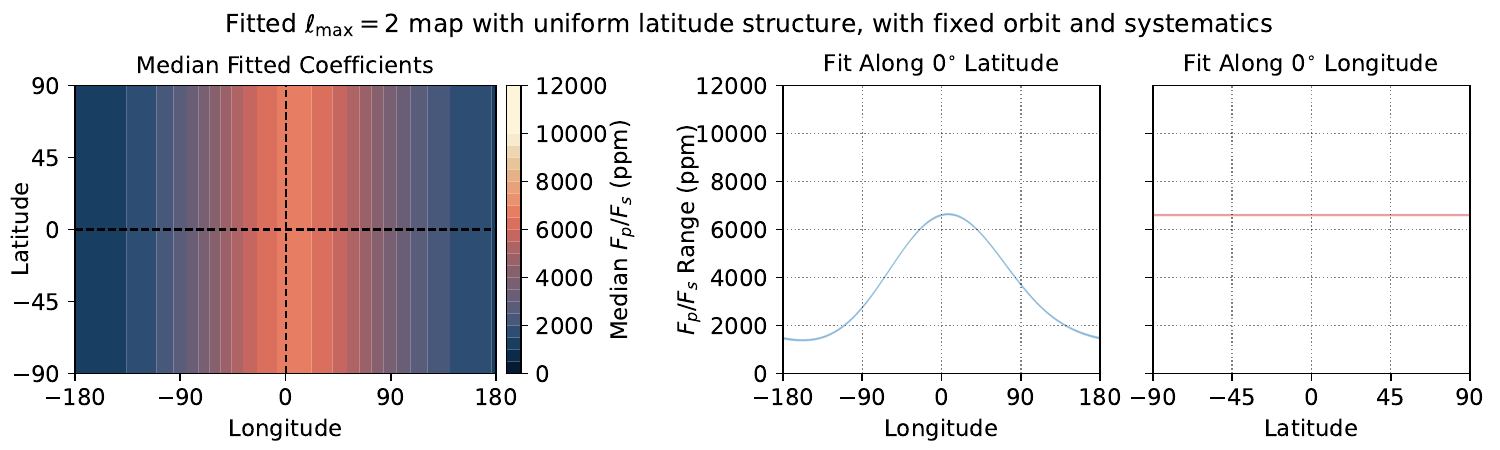}
    \caption{An eclipse map fitted to the dataset in Figure \ref{fig:fig1_lightcurve}, using the parameters listed in Table \ref{tab:params_table} derived with the $\ell_{\rm max}=2$ eclipse map model, and constructed from $\ell_{\rm max}=2$ spherical harmonics averaged in latitude. This is distinct from the ``Fourier series'' phase curve model used in Table \ref{tab:params_table}, which models the phase curve directly and does not represent the partial coverage of the map in eclipse. The fit to the ingress and egress data is shown in red in Figure \ref{fig:fig2_residual}; as described in Section \ref{sec:result:signal}, this map cannot accurately fit the observed eclipse shape, showing the presence of latitudinal information in the observations.}
    \label{fig:fig4_o2_flat_map}
\end{figure*}

\section{Results} \label{sec:results}

\subsection{Eclipse map fitted with orbit and systematics}\label{sec:results:map}

The red line in Figure \ref{fig:fig1_lightcurve} shows the phase curve resulting from the $\ell_{\rm max}=2$ eclipse map fitted alongside the orbital parameters, stellar parameters, and systematic model. The blue line in Figure \ref{fig:fig1_lightcurve} shows the model of instrumental systematics fitted at the same time. We normalised the $F_{p}/F_{S}$ values (the ratio of planetary to stellar flux) in Figure \ref{fig:fig1_lightcurve} so that the systematic model has zero mean over the course of the fitted observations. 

Appendix \ref{ap:params} shows the posterior distribution for the fitted orbital, planetary, stellar, and systematic parameters. Some of the parameters are consistent with those derived using a Fourier series fit like that used in \citet{bell2023wasp43b}, but some are not consistent. There are small but statistically significant differences in the planetary radius, eclipse timing, and inclination, all of which we expect to be more accurately fitted by the more realistic eclipse shape in the eclipse mapping model. The posterior distribution of the stellar limb darkening parameters are different but we found that, when combined, these resulted in consistent limb darkening profiles. The stellar limb darkening is relatively weak at these long wavelengths, so is poorly constrained but is consistent between the two models. The systematic parameters are different to those produced by a Fourier series fit \citep{bell2023wasp43b} but result in an almost identical model of systematics, as a short exponential ramp plus a linear trend is almost identical to a long exponential ramp.

Figure \ref{fig:fig1_lightcurve} shows that this $\ell_{\rm max}=2$ eclipse map model fits the observed phase curve well. The information about the 2D eclipse map structure is contained within the shapes of the ingress and egress of each eclipse. Figure \ref{fig:fig2_residual} shows the residual difference in ingress and egress between the observations (shown as black, green, and purple points), and a ``control'' model representing the phase curve as a Fourier series up to order $n=2$. This Fourier series was fitted with orbital and systematic parameters fixed to the values derived when fitting the $\ell_{\rm max}=2$ eclipse map, listed in Table \ref{tab:params_table}, so that the residual is only due to the different emission models. 

The green and purple points in Figure \ref{fig:fig2_residual} show the residual of the observed data in the first and second eclipses, binned every 8 points, and the black points show the average of the two eclipses. The $\ell_{\rm max}=2$ eclipse mapping model and the $n=2$ Fourier series model fit the out-of-eclipse phase curve as well as each other, so the residual outside the range plotted in Figure \ref{fig:fig2_residual} is determined by the uncertainty in the data. This is not surprising, as both models have access to almost identical fitting functions for the out-of-eclipse phase curve -- the $\ell_{\rm max}=2$ harmonics produce the same phase curves as the $n=2$ Fourier series, apart from a small modification from the orbital obliquity.

The size of the residual between the observed data and the $n=2$ control model shows the size of the eclipse mapping signal, which is the effect of partial stellar coverage of non-uniform emission from the planet \citep{de2012towards}. For example, when the star covers an area that is emitting more than the average of the planetary disk, the observer measures less flux than would be expected for a uniform disk, so the residual signal is negative. 

The blue shaded area shows the range of phase curves fitted using the $\ell_{\rm max}=2$ eclipse map model, with the Fourier series control model subtracted. The two shaded regions show the first and second quantiles, containing $68.27\%$ and $95.45\%$ of the posterior distribution respectively. The eclipse map model matches the residual signal in ingress and egress, with a uncertainty comparable to the error bars of the observed data. This implies a good fit to a robust eclipse mapping signal, which we quantify in the next section.

Figure \ref{fig:fig3_o2_map} shows the $\ell_{\rm max}=2$ eclipse map itself. The plotted 2D map is constructed from the median values of the posterior distribution of each of the spherical harmonic coefficients. This median map has a hot-spot near the substellar point, with a small meridionally averaged shift of $(7.75^{+0.36}_{-0.36})^{\circ}$ eastwards and a shift of $( 10.72 ^{+ 4.14 } _{- 4.68 })^{\circ}$ southwards\footnote{The observations cannot distinguish between an orbit about a vector pointed ``up'' or ``down'' in the sky, so our choice of north and south poles is made arbitrarily and kept consistent for plotting. The eastward direction corresponds to the direction of rotation of the planet.}.

We present the maps as a ratio of planetary flux to stellar flux, as a conversion to temperature is more complex and requires modelling or assumptions about the temperature structure. We make a simple estimate of the brightness temperature assuming planetary black-body emission $B_{T,\lambda}$ in the bandpass $\lambda=5$ to 10.5 $\mu$m, and a PHOENIX \citep{allard1995, hauschildt1999, husser2013} stellar model spectrum with effective temperature $T_{\rm eff}$ = 4300~K and surface gravity $\log g$ = 4.50 as used in \citet{bell2023wasp43b}. The stellar radius that we assumed to be 0.665\,R$_{\odot}$ does affect the value of the brightness temperature. We do not include effect of the uncertainty of this and other stellar parameters on the derived brightness temperature, in order to highlight the uncertainty due to mapping alone. This determines that this $\ell_{\rm max}=2$ map corresponds to a brightness temperature of $( 1790.0 ^{+ 23.0 } _{- 29.0 }) $ K at the substellar point, $( 1293.0 ^{+ 27.0 } _{- 34.0 }) $ K at the equatorial east terminator, $( 1114.0 ^{+ 30.0 } _{- 36.0 }) $ K at the equatorial west terminator, $( 1100.0 ^{+ 128.0 } _{- 109.0 }) $ K at the south pole, and $( 1011.0 ^{+ 111.0 } _{- 108.0 }) $ K at the north pole (where the distinction between the poles is our arbitrary choice).

The posterior distribution of the fitted maps is shown east-west along the equator and north-south through the substellar point. This range of fitted maps includes any degeneracies with the orbital or systematic parameters as these are fitted simultaneously. We confirmed that the ``median map'' plotted in 2D corresponds closely to the center of the east-west and north-south posterior distributions (not shown). The maximum likelihood map is very similar to the median map, except near the poles where its latitudinal structure deviates slightly from the median but remains within the first quantile.

Figure \ref{fig:fig4_o2_flat_map} shows an $\ell_{\rm max}=2$ eclipse map fitted to the same dataset, but with all of its fitting basis maps averaged in latitude, removing the ability of the map to fit latitudinal structure. This makes some of the basis maps uniform everywhere, so we remove them from the fitting process (reducing the number of parameters). The resulting map is very tightly constrained by its need to fit the out-of-eclipse phase curve. Its residual signal in ingress and egress is shown by the red region in Figure \ref{fig:fig2_residual}, which fails to match the observed residual points. This shows the presence of a latitudinal mapping signal of around 200 ppm (the difference between this longitude-only fit, and the observed residual), compared to the longitudinal mapping signal of around 250 ppm (the magnitude of the residual of this longitude-only fit).

\begin{figure*}
    \centering
    \includegraphics[width=\textwidth]{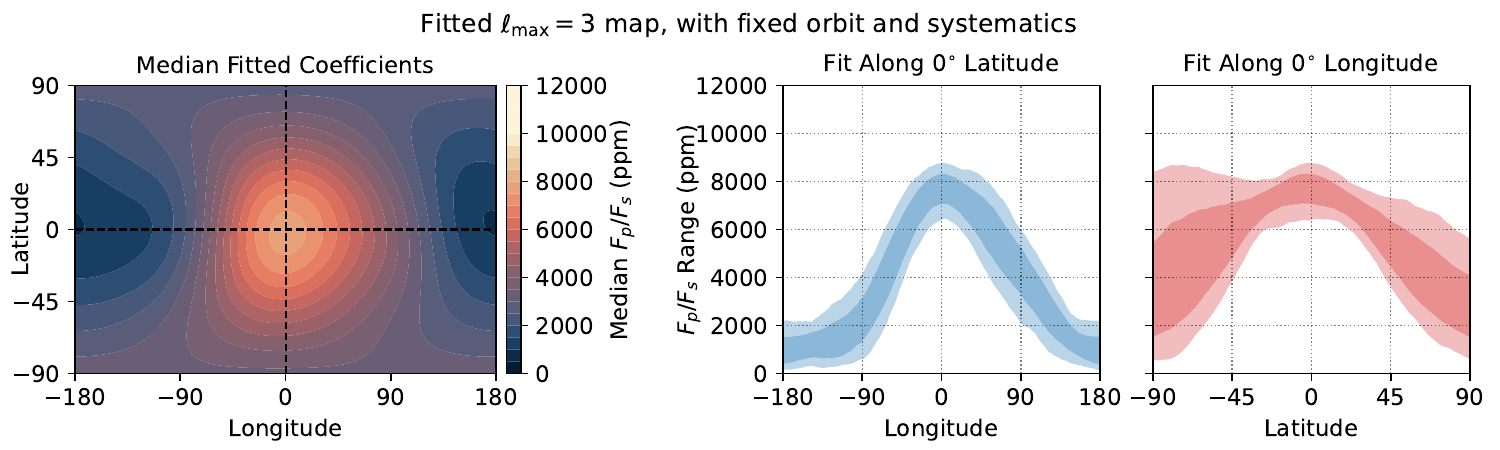}
    \includegraphics[width=\textwidth]{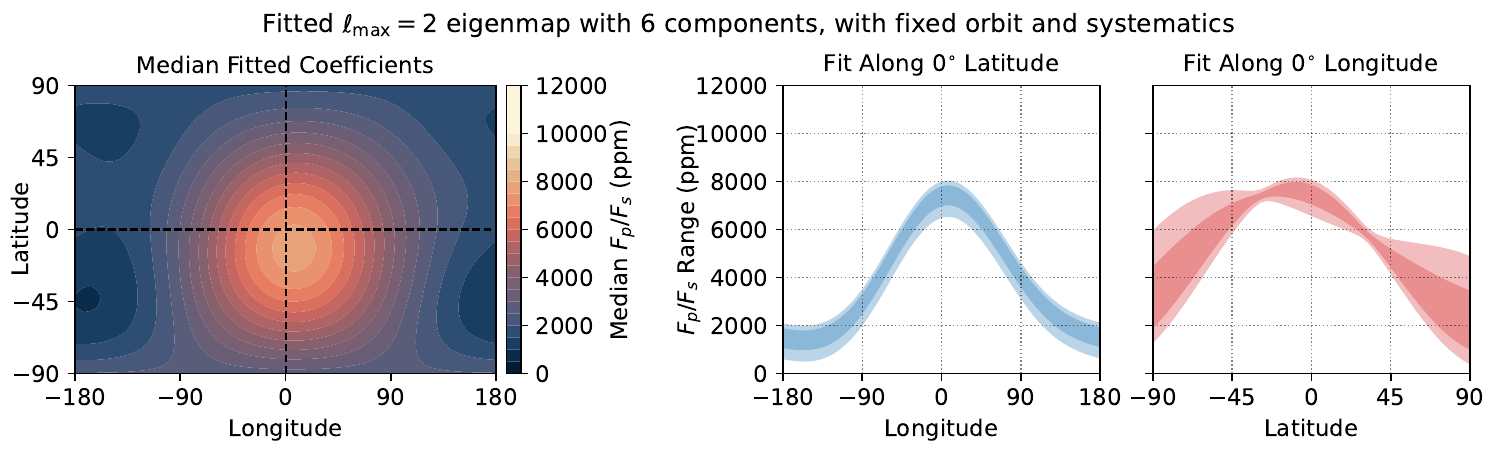}
    \caption{Two alternative mapping methods to Figure \ref{fig:fig3_o2_map}. First row: an eclipse map using spherical harmonics up to order $\ell_{\rm max}=3$, fitted using constant systematic and orbital parameters derived by the map fitted in Figure \ref{fig:fig3_o2_map} (listed in Table \ref{tab:params_table}) The median map, longitudinal structure, and latitudinal structure are laid out as in Figure \ref{fig:fig3_o2_map}. Bottom row: an eclipse map fitted with eigenmaps up to order $\ell_{\rm max}=3$ as described in Section \ref{sec:results:eigenmap}, for six basis maps selected by optimising the BIC, using constant systematic and orbital parameters derived by the fit in Figure \ref{fig:fig3_o2_map}.}
    \label{fig:fig5_o3_map_eigenmap}
\end{figure*}

\subsection{Eclipse Mapping Signal}\label{sec:result:signal}

We can quantify the significance of the signal in Figure \ref{fig:fig2_residual} by comparing the Bayesian evidence of the $\ell_{\rm max}=2$ eclipse map model (the blue shaded region) and the $n=2$ Fourier series control model (the zero line) in Figure \ref{fig:fig2_residual}. As applied in \citet{placek2017analyzing} to a similar analysis of exoplanet phase curves, the log-likelihood function is:
\begin{equation}
    \log L=-\frac{1}{2 \sigma^2} \sum_{i=1}^N\left(F(t_{i})-D_{i}\right)^2-\frac{N}{2} \log 2 \pi \sigma^2,
\end{equation}
for data $D_{i}$ at $N$ times $t_{i}$ fitted by a model $F(t)$ with variance $\sigma^{2}$ . The log-odds ratio $\ln O = \ln L_{1} - \ln L_{2}$ then represents the relative performance of models 1 and 2, with a value of $\ln O$ above 1 corresponding to a better fit to model 1. The log-odds ratio for the $\ell_{\rm max}=2$ map model compared to the Fourier series model (with the eclipse shape of a uniformly emitting disk) is 52.6, showing that the eclipse model is overwhelmingly preferred according to the criterion in \citet{placek2017analyzing}. Both models fit the out-of-eclipse data very similarly as they have access to almost identical out-of-eclipse fitting functions, so this difference in likelihood is entirely due to the improved fit in the ingress and egress of the eclipses shown in Figure \ref{fig:fig2_residual}. Later, we use the Bayesian Information Criterion (BIC) to compare the fit quality of the eclipse map models while taking the number of fitted parameters into account. We cannot use the BIC to compare a map model to the Fourier series model as they are not nested.

Notably, the log-odds ratio of an $n=2$ Fourier series fit with free orbital, planetary, stellar, and systematic parameters (used to derive the parameters in Table \ref{tab:params_table}) is 20.7 compared to the $\ell_{\rm max}=2$ map model. This shows how the non-mapping parameters are modified by the Fourier series model to slightly better fit the ingress and egress shape. However, this makes a minor difference overall and the eclipse map model is still very strongly preferred.

Comparing the eclipse map and Fourier series models confirms that there is a strong eclipse mapping signal in this dataset. This does not immediately imply the presence of 2D information in the dataset, as \citet{coulombe2023wasp18b} only found evidence for longitudinal information in their eclipse map of WASP-18b. We can show that latitudinal information is present by comparing the 2D $\ell_{\rm max}=2$ map to the ``1D'' $\ell_{\rm max}=2$ map where all of the basis maps are averaged in latitude, shown in Figure \ref{fig:fig4_o2_flat_map}. The red region in Figure \ref{fig:fig2_residual} shows the residual eclipse mapping signal for this 1D map, which does not fit the real dataset as well as the 2D map.  

The log-odds ratio of the 2D map to the 1D map fits in Figure \ref{fig:fig2_residual} is 9.8, showing that the 2D map with variable latitudinal structure is greatly preferred.  Notably, the uncertainty on the residual of the 1D map in Figure \ref{fig:fig2_residual} is very small, as the longitudinal structure is well constrained by the additional information from the out-of-eclipse phase curve. This implies that most of the uncertainty on the 2D model in Figure \ref{fig:fig2_residual} is related to uncertainty about the latitudinal structure of the map.

\citet{coulombe2023wasp18b} conducted the same comparison between 1D and 2D eclipse maps of WASP-18b and found no evidence that the 2D map was preferred over the 1D map. They therefore concluded that the eclipse mapping signal revealed longitudinal information only. In contrast, the eclipse mapping signal in this study of WASP-43b is sensitive to both longitudinal and latitudinal information, as would be expected from the higher impact parameter of the orbit. This is therefore the first time that latitudinal information has been shown to be detected on an exoplanet.

\subsection{Eclipse map fitted with $\ell_{\rm max}=3$ spherical harmonics}\label{sec:results:map_o3}

The spherical harmonic order of the $\ell_{\rm max}=2$ map fit to the dataset was limited by the number of parameters that could be fitted simultaneously with the orbital, stellar, and systematic parameters. Figure \ref{fig:fig10_corner} in Appendix \ref{ap:params} shows the posterior distribution for these parameters, demonstrating that there are no significant degeneracies. We therefore fix these parameters to the values (listed in Table \ref{tab:params_table}) derived using the $\ell_{\rm max}=2$ eclipse map model and re-fit the map with a variety of methods.

\citet{bell2023wasp43b} found that an $n=1$ Fourier series fitted the out-of-eclipse phase curve very poorly, so we do not consider any $\ell_{\rm max}=1$ spherical harmonic fits. The top row of Figure \ref{fig:fig5_o3_map_eigenmap} shows the eclipse map re-fitted using spherical harmonics up to order $\ell_{\rm max}=3$.  We fix the orbital, planetary, stellar, and systematic parameters to the median values derived from the fit in Figure \ref{fig:fig3_o2_map}. We found that the $\ell_{\rm max}=3$ maps required too many pixels to sample effectively with \textit{PyMC3}, so in this case we sampled the spherical harmonic coefficients directly with \textit{mc3} \citep{cubillos2016correlated}. This method allows us to impose positivity by excluding negative maps from the fitting process, instead of sampling pixels with positive priors. We set the spherical harmonic coefficients to have Gaussian priors with a mean of 0 and a standard deviation equal to the mean of the observed planetary flux.

The resulting map in the top row of Figure \ref{fig:fig5_o3_map_eigenmap} shows a similar median structure to the map in Figure \ref{fig:fig3_o2_map}, with more uncertainty due to the increased degrees of freedom. There is a longitudinal offset of $(0.50^{+14.79}_{-8.04})^{\circ}$, which  is consistent with the $\ell_{\rm max}=2$ fit value of $(7.75^{+0.36}_{-0.36})^{\circ}$. Its median value is almost zero, which may seem unusual as the overall phase curve has an offset of $(7.3^{+0.4}_{-0.4})^{\circ}$. This apparent discrepancy is due to the more complex shapes allowed by the $\ell_{\rm max}=3$ spherical harmonics, which produce a dayside hot-spot which peaks at the substellar point but extends further east than west (as shown in Figure \ref{fig:fig5_o3_map_eigenmap}) resulting in an overall shift in the phase curve. We discuss this distinction between hot-spot position and phase curve offset in Section \ref{sec:discussion}, showing how the phase curve offsets and hot-spot positions can differ in numerical simulations. 

The latitudinal structure of the $\ell_{\rm max}=3$ map is somewhat different to the $\ell_{\rm max}=2$ map, being flatter near the equator with no significant latitudinal hot-spot shift. We will also discuss in Section \ref{sec:discussion} how latitudinal structure could be degraded by using low-order basis maps to fit the data. For all these reasons, the spatial resolution of the map used to fit the data is a key question. In Section \ref{sec:results:model_selection}, we will therefore discuss how many parameters  are justified to fit the dataset.

\subsection{Eclipse map fitted with eigenmapping}\label{sec:results:eigenmap}

The bottom row of Figure \ref{fig:fig5_o3_map_eigenmap} shows a map fitted using ``eigenmapping'' \citep{rauscher2018eigenmap} with ThERESA \citep{challener2022theresa}, with the orbital and systematic parameters again fixed to the median values derived from the $\ell_{\rm max}=2$ fit in Figures \ref{fig:fig1_lightcurve} and \ref{fig:fig3_o2_map}.

Briefly, eigenmapping starts with spherical harmonic phase curves up to some degree $\ell_{\rm max}$, orthogonalizes them with principal component analysis to create ``eigencurves'', and fits the observed phase curve as a sum of $N_E$ highest-variance eigencurves. $\ell_{\rm max}$ and $N_E$ are selected to minimize the BIC. Each eigencurve has a corresponding eigenmap, and the fitted map is the corresponding sum of these eigenmaps. We again enforce a positive-flux constraint on the fitted map globally. Eigenmapping has been used to map  observations of HD 189733b with the \textit{Spitzer Space Telescope} \citep{rauscher2018eigenmap, challener2022theresa}, and observations of WASP-18b with JWST NIRISS/SOSS \citep{coulombe2023wasp18b}. 

We find the best BIC for $\ell_{\rm max} = 2$ and $N_E = 6$, a larger number of eigencurves than used for WASP-18b in \citet{coulombe2023wasp18b}, showing the improved precision of this WASP-43b dataset. We find a slight eastward hot-spot offset of $(7.5^{+0.5}_{-0.5})^{\circ}$. This eigenmap achieves the best BIC score in Table \ref{tab:model_comparison}; its fitted map is almost the same as the $\ell_{\rm max}=2$ map in Figure \ref{fig:fig3_o2_map}, suggesting that it achieves this improved BIC score by discarding the map components that contribute the least to the observed phase curve. This means that it produces the same fit quality using fewer parameters (see Section \ref{sec:discussion} for a related discussion of the mapping ``null space'').

\begin{figure*}
    \centering
    \includegraphics[width=\textwidth]{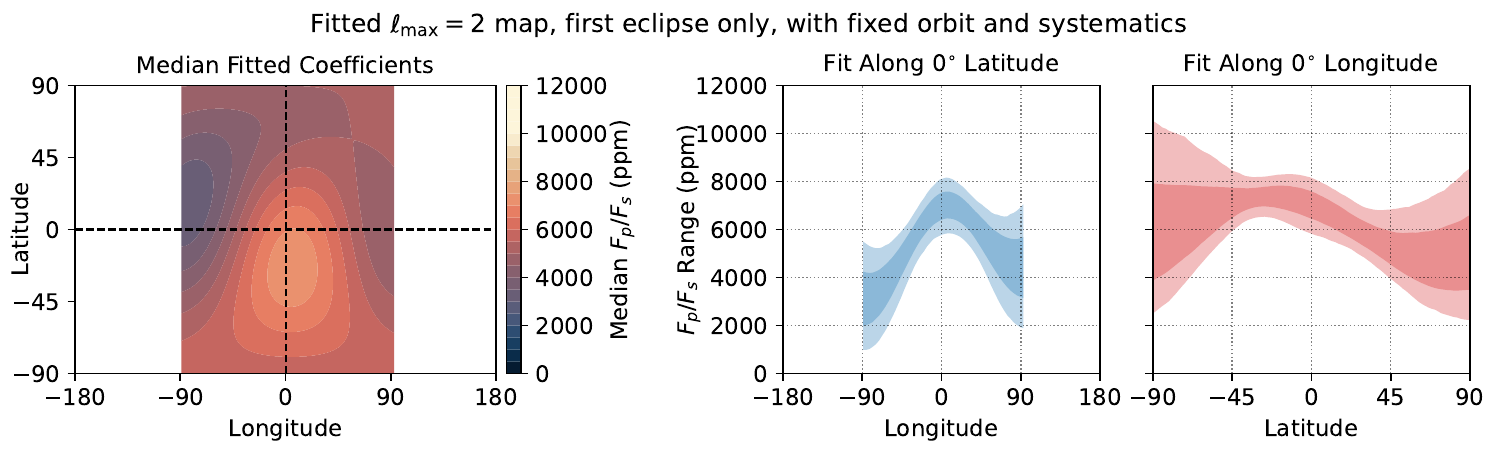}
    \includegraphics[width=\textwidth]{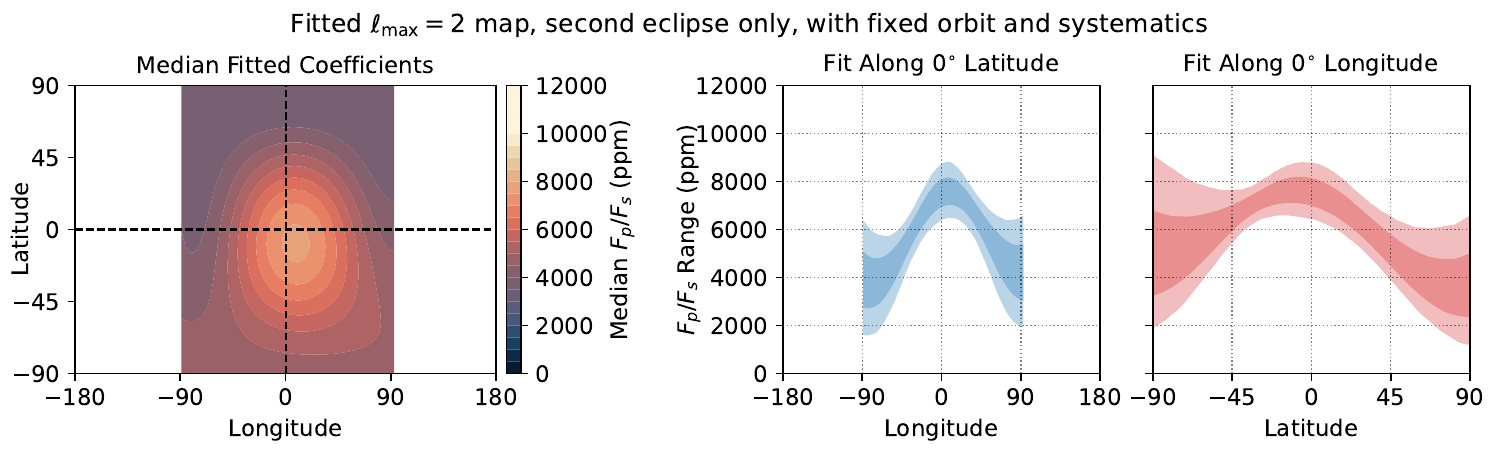}
    \includegraphics[width=\textwidth]{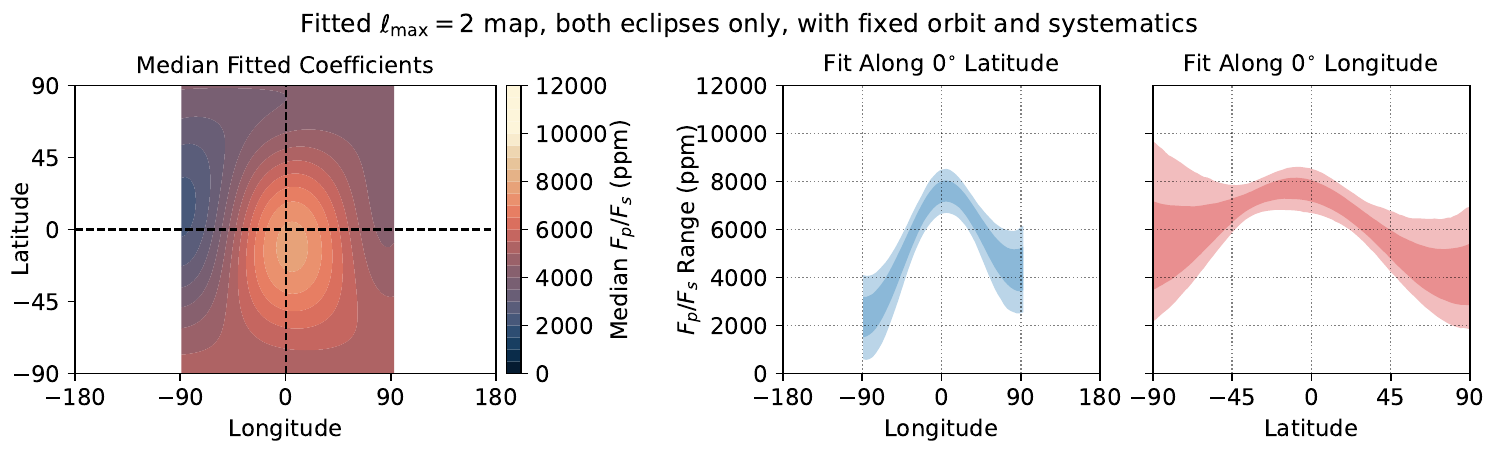}
    \caption{Eclipse maps fitted using each eclipse separately. First row: an eclipse map fitted using spherical harmonics up to $\ell_{\rm max}=2$, using a $\sim2.2$ hour section of the data centered on the first eclipse. The orbital and systematic parameters are fixed to those derived using the map fitted in Figure \ref{fig:fig3_o2_map} (except the time-dependent systematics, which are re-fitted), listed in Table \ref{tab:params_table}. Only the dayside of the map is shown as there is no information about the nightside in this limited dataset. The map has a similar longitudinal structure to Figure \ref{fig:fig3_o2_map}, but finds more latitudinal asymmetry. Second row: an eclipse map fitted using a $\sim2.2$ hour section of the data centered on the second eclipse. This finds a similar map to that in Figure \ref{fig:fig3_o2_map}. These two maps show that both eclipses are mostly consistent, although the first eclipse implies more latitudinal asymmetry. Third row: an eclipse map fitted using both the first and second eclipses as defined above, excluding the rest of the phase curve. This combined fit is similar to the individual fits, especially in its longitudinal structure.}
    \label{fig:fig6_e1_e2_ee_map}
\end{figure*}

\subsection{Eclipse maps fitted with single eclipses}\label{sec:results:single}

The maps in Figures \ref{fig:fig3_o2_map} and \ref{fig:fig5_o3_map_eigenmap} are fitted to the entire dataset, including 2D information from both eclipses as well as 1D information from the rest of the phase curve. To isolate the 2D mapping information in each eclipse, Figure \ref{fig:fig6_e1_e2_ee_map} shows the result of re-fitting the $\ell_{\rm max}=2$ eclipse map using the first and second eclipses only. As in Figure \ref{fig:fig5_o3_map_eigenmap}, we fix the orbital and systematic parameters to the median values derived with the model in Figures \ref{fig:fig3_o2_map}, listed in Table \ref{tab:params_table}. The exceptions are the magnitude of the exponential ramp and the magnitude of the linear trend, which we re-fit as these require different values given the different starting points in time

The first row in Figure \ref{fig:fig6_e1_e2_ee_map} shows an $\ell_{\rm max}=2$ eclipse map fitted using a $\sim2.2$ hour section of data containing the first eclipse only with a small section of phase curve either side. We only show the map on the dayside of the planet, as there is no information about the nightside contained in this limited dataset. The map is similar to Figure \ref{fig:fig3_o2_map} and \ref{fig:fig5_o3_map_eigenmap}, but has a larger latitudinal hot-spot shift than the map in Figure \ref{fig:fig3_o2_map}. The latitudinal peak of the fit to the first eclipse only is weakly constrained, showing how an observation of only a single eclipse is likely to be insufficient to derive an accurate eclipse map.  There is a greater range of fitted maps at $-90^{\circ}$ latitude than $+90^{\circ}$ latitude because the inclination of the orbit angles this pole away from the observer.

The second row in Figure \ref{fig:fig6_e1_e2_ee_map} shows an eclipse map fitted to the second eclipse only, with a smaller latitudinal hot-spot shift than the fit to the first eclipse only. The third row in Figure \ref{fig:fig6_e1_e2_ee_map} shows an eclipse map fitted to both eclipses together. We suggest that the original fit to the entire dataset in Figure \ref{fig:fig3_o2_map} which found a latitudinal shift of $( -10.72 ^{+ 4.14 } _{- 4.68 })^{\circ}$ is better constrained in latitude than this two-eclipse fit (despite containing no additional latitudinal information) because the additional measurement of the rest of the phase curve provides independent longitudinal information. This breaks degeneracies between the longitudinal and latitudinal information in the eclipse mapping signal \citep{boone2023analytic}. 

The maps in Figure \ref{fig:fig6_e1_e2_ee_map} show that the eclipse mapping signals in both eclipses are mostly consistent with each other, as would be expected from the similar residual signals in Figure \ref{fig:fig2_residual}. However, they have different degrees of latitudinal asymmetry, which also manifests in the overall fit in Figure \ref{fig:fig3_o2_map}. Figure \ref{fig:fig6_e1_e2_ee_map} shows that the first eclipse implies more latitudinal asymmetry. This may be a real effect, but it may also be due to the increased effect of instrumental systematics on this eclipse. The blue line in Figure \ref{fig:fig1_lightcurve} shows that the systematic signal is still relatively large during the first eclipse, but that it is almost negligible during the second eclipse. This shows the utility of observing a full phase curve where the periodicity of the astrophysical signal isolates it from the systematic signal.

\subsection{Model Selection}\label{sec:results:model_selection}

To compare our fitted eclipse maps, we calculate the Bayesian Information Criterion (BIC) \citep{schwarz1978estimating} and the Akaike Information Criterion (AIC) \citep{akaike1981likelihood}. The BIC is:

\begin{equation}
    \mathrm{BIC}=k \ln (N)- 2 \ln (\widehat{L}),
\end{equation}

where $k$ is the number of model parameters, $N$ is the number of data points, and $\widehat{L}$ is the model likelihood:

\begin{equation}
    \ln (\widehat{L}) = -\frac{1}{2} \sum_{i}^{N} \left( \frac{M_{i} - D_{i}}{\sigma_{i}} \right)^{2},
\end{equation}

for a sum over all $N$ data points $D_{i}$ with uncertainty $\sigma_{i}$, fitted by a model $M_{i}$. This uncertainty is derived from the residuals of the originally fitted phase curve, by scaling the expected error by a factor referred to as the ``Uncertainty Scaling Factor'' in Figure \ref{fig:fig10_corner} (see \citet{bell2023wasp43b}).

A smaller BIC implies a better model, with a better fit or fewer fitting parameters. The difference in BIC is the relevant quantity for model comparison, so Table \ref{tab:model_comparison} shows the $\Delta$BIC for each model compared to Model M5, the eigenmap in Figure \ref{fig:fig5_o3_map_eigenmap}. This criterion allows comparison of nested models where it is assumed that the true model is inside the set of tested models \citep{burnham2004multimodel}. 

We also consider the AIC of each model, which is:

\begin{equation}
    \mathrm{AIC}= 2 k - 2 \ln (\widehat{L}),
\end{equation}

Due to the large number of points $N$ in this dataset, the BIC applies a stronger penalty for each parameter than the AIC. We also present the ``weights'' of each model, \textit{w}BIC and \textit{w}AIC, which are the relative probabilities of each compared model in a set, based on their $\Delta$BIC or $\Delta$AIC. The \textit{w}BIC for a particular model is:

\begin{equation}
    w\mathrm{BIC}=\frac{\exp \left(-\Delta\mathrm{BIC} / 2\right)}{\sum_{r=1}^R \exp \left(-\Delta\mathrm{BIC}_{r} / 2\right)}
\end{equation}

where $\Delta$BIC is compared to the best-performing model, and there are $R$ total models being compared with a score of $\Delta\mathrm{BIC}_{r}$ each. \textit{w}AIC depends on $\Delta$AIC in the same way. Table \ref{tab:model_comparison} summarises our compared models, showing their $\chi^{2}$ values, their $\Delta$BIC and $\Delta$AIC relative to the model with the best BIC and AIC, and the implied probability of each model.

Model M1 is the Fourier series model used in \citet{bell2023wasp43b}. It achieves a significantly worse $\chi^{2}$ value than all the eclipse mapping models, due to a worse fit to the eclipse shape. Model M2 is the eclipse map model shown in Figure \ref{fig:fig4_o2_flat_map}, where the latitudinal structure is fixed to be flat. This achieves a better $\chi^{2}$ value than the Fourier series model, as it includes the effect of longitudinal map structure on the eclipse shape. However, its $\chi^{2}$ value is worse than the fully two-dimensional eclipse map fits, as shown by its poor fit in Figure \ref{fig:fig2_residual}. Interestingly, due to its smaller number of parameters, it achieves the second-best BIC score of all the eclipse mapping models due to the strong penalty applied to the number of parameters by the BIC. However, it achieves the worst AIC score of the eclipse map models, where the number of parameters is penalised less strongly. We can confidently reject models M1 and M2 compared to the 2D eclipse mapping models.

Model M3 is the $\ell_{\rm max}=2$ eclipse mapping model in Figure \ref{fig:fig3_o2_map}. It achieves a better $\chi^{2}$ value than models M1 and M2, as shown by its good fit in Figure \ref{fig:fig2_residual}. It has a worse BIC score than the eigenmap (model M5), as they have similar $\chi^{2}$ values, but model M3 uses two more parameters. It has a more similar AIC score to the eigenmap, where these additional parameters are penalized less heavily.

Model M4 is the $\ell_{\rm max}=3$ eclipse mapping model in Figure \ref{fig:fig5_o3_map_eigenmap}. It achieves the best $\chi^{2}$ value due to its increased degrees of mapping freedom, but has poor BIC and AIC scores due to its increased number of parameters. These metrics imply that the data quality does not justify this number of degrees of freedom, which is also implied by the large uncertainty on the $\ell_{\rm max}=3$ map in Figure \ref{fig:fig5_o3_map_eigenmap}.

Model M5 is the $\ell_{\rm max}=2$ eigenmap using 6 basis maps. It achieves a similar $\chi^{2}$ to the other eclipse mapping models, and has the best BIC and AIC scores due to its reduced number of parameters. It achieves these by discarding mapping structures that contribute weakly to the observed phase curve, so is able to match the observations with fewer degrees of freedom.

It is not clear whether the BIC or AIC is the better metric for model comparison, or if another metric would be more appropriate. The BIC assumes that the set of fitted models includes the ``true'' physical system, while the AIC assumes that all of the tested models are inexact representations of this system \citep{burnham2004multimodel}. It could be argued that eclipse maps fitted to real planets do include the ``true'' map, as spherical harmonics form a complete basis set on the sphere. On the other hand, information about the real map is inevitably lost in the mapping process due to the need to truncate fits to low-order harmonics, and due to the presence of a ``null space'' (see Section \ref{sec:discussion} for discussion of both issues). 

To summarise our model comparison,

\begin{enumerate}
    \item The Fourier series model (M1), and the eclipse map with flat latitudinal structure (M2) perform badly on all statistical metrics, so we confidently reject them.
    \item The latitude-longitude $\ell_{\rm max}=2$ eclipse map (M3) is favoured over the longitude-only eclipse map (M2), showing the presence of latitudinal information in the data.
    \item The $\ell_{\rm max}=3$ eclipse map achieves a better $\chi^{2}$ value than model M3, but at the cost of many more parameters, so has worse BIC and AIC scores.
    \item The $\ell_{\rm max}=2$ eigenmap achieves the best BIC and AIC scores, producing a fit of comparable quality to model M3 with two fewer parameters
\end{enumerate}

Therefore, the eigenmap (model M5) is the best performing model on these metrics. However, it requires model M3 to fit the orbital, stellar, and systematic parameters simultaneously first, and then both models produce a similar result in the end. The eigenmap achieves a better AIC and BIC by discarding structures that do not contribute strongly to the map, producing a very similar map to model M3 overall. We suggest that more work can be done on the process of fitting eclipse maps and comparing levels of model complexity. Cross-validation may provide a more practical metric for eclipse mapping model comparison, such as the leave-one-out cross-validation applied by \citet{challener2023bringing}, which is asymptotically equivalent to the AIC but provides advantages such as an absolute measurement of model predictive power \citep{stone1977asymptotic}.

\begin{table*}
\centering
\begin{tabular}{ll|ccccccc}
\textbf{Model}   & \textbf{Order} & Longitude Offset & Parameters & $\chi^{2}$ & $\Delta$BIC & $\Delta$AIC & \textit{w}BIC & \textit{w}AIC \\
\hline
M1: Fourier Series & $n=2$   & $(7.3^{+0.4}_{-0.4})^{\circ}$    & 5  & 8121.0  & --  & -- & -- & -- \\
M2: Uniform Latitude &  $\ell_{\rm max}=2$ &  $(7.3^{+0.4}_{-0.4})^{\circ}$   & 5 & 8039.9 & 6.4  & 20.4 & $3.9\%$ & $0.0\%$  \\
M3: Spherical Harmonics &  $\ell_{\rm max}=2$ &  $(7.75^{+0.36}_{-0.36})^{\circ}$   & 10 & 8016.5 &  19.0 & 5.0 & $0.0\%$ & $7.5\%$  \\
M4: Spherical Harmonics & $\ell_{\rm max}=3$ &  $(0.50^{+14.79}_{-8.04})^{\circ}$  & 17 & 8014.9 & 80.2 & 17.4 & $0.0\%$ & $0.0\%$  \\
M5: Eigenmap & $\ell_{\rm max}=2$, $N_{E}=6$       &  $(7.5^{+0.5}_{-0.5})^{\circ}$  & 8 & 8015.5  & 0 & 0 & $96.1\%$ & $92.5\%$   \\ 
\end{tabular}
\caption{A comparison of the models used to fit the full dataset. The offset value for the Fourier Series model is the offset of the phase curve, and the offset value for the eclipse maps is the meridionally averaged longitudinal offset as defined in Section \ref{sec:discussion:offsets}. The $\Delta$BIC and $\Delta$AIC scores are calculated relative to the eigenmap model. When calculating the $\chi^{2}$, BIC, and AIC values, we exclude the transit because ThERESA \citep{challener2022theresa} does not model this explicitly.  We do not calculate a comparative BIC or AIC between the Fourier series model and the eclipse map models as they are not nested.}\label{tab:model_comparison}
\end{table*}

\section{Comparison to General Circulation Models}\label{sec:discussion}

In this section, we interpret the eclipse maps by comparing them to three-dimensional General Circulation Models (GCMs). We use some of the GCMs presented in \citet{bell2023wasp43b}, selecting those that matched the out-of-eclipse phase curve well. We include a simulation from four models: \emph{THOR} (\citealp{mendoncca2016thor,mendoncca2018revisiting,mendoncca2018three}; Simulation 31 in \citealp{bell2023wasp43b}), \emph{expeRT/MITgcm} (\citealp{carone2020equatorial,schneider2022exploring}; Simulation 20 in \citealp{bell2023wasp43b}), the Generic Planetary Climate Model (\emph{PCM}) (\citealp{teinturier2023}; Simulation 7 in \citealp{bell2023wasp43b}), and the \emph{RM-GCM} (\citealp{rauscher2012general,roman2017modeling,roman2021clouds}; Simulation 24 in \citealp{bell2023wasp43b}). 

The THOR and RM-GCM simulations use semi-grey radiative transfer, while the PCM and expeRT/MITgcm simulations use multi-band correlated-k schemes. The THOR, RM-GCM, and PCM simulations feature clouds (on the nightside only for THOR), whereas the expeRT/MITgcm simulation is free of clouds. Output from each model was post-processed using several multi-band radiative transfer codes to calculate spectrally resolved emission from each column, which was then integrated from $5$ to $10.5\,\mu\text{m}$, weighted by the MIRI/LRS throughput. Appendix \ref{ap:gcms} and \citet{bell2023wasp43b} give more detail on each model.

\subsection{Observable Features of GCMs}\label{sec:discussion:gcms}

\begin{figure*}[!t]
    \centering
    \includegraphics[width=\textwidth]{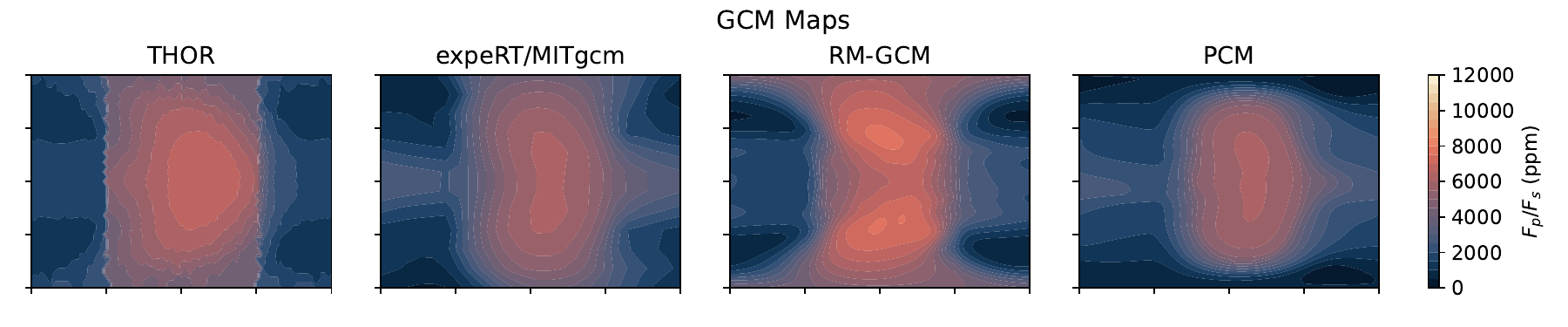}
    \includegraphics[width=\textwidth]{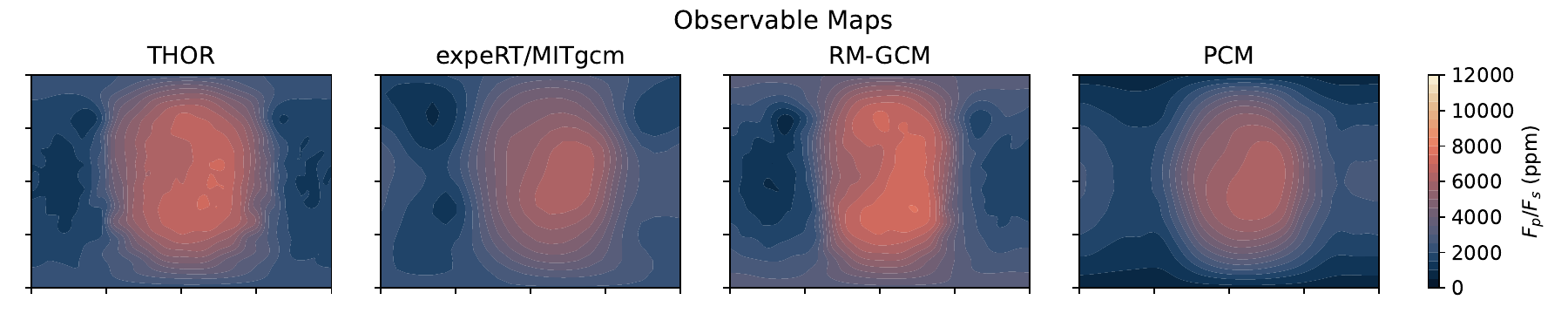}
    \includegraphics[width=\textwidth]{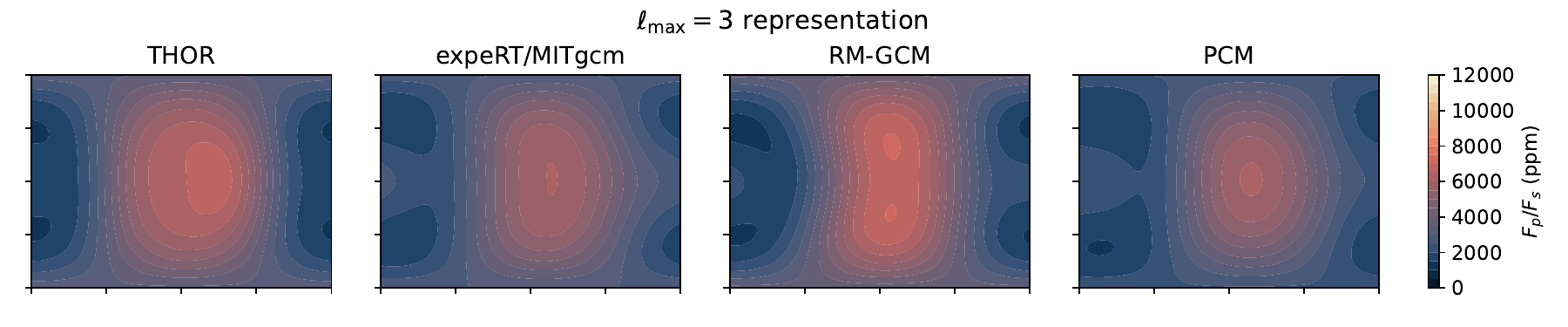}
    \includegraphics[width=\textwidth]{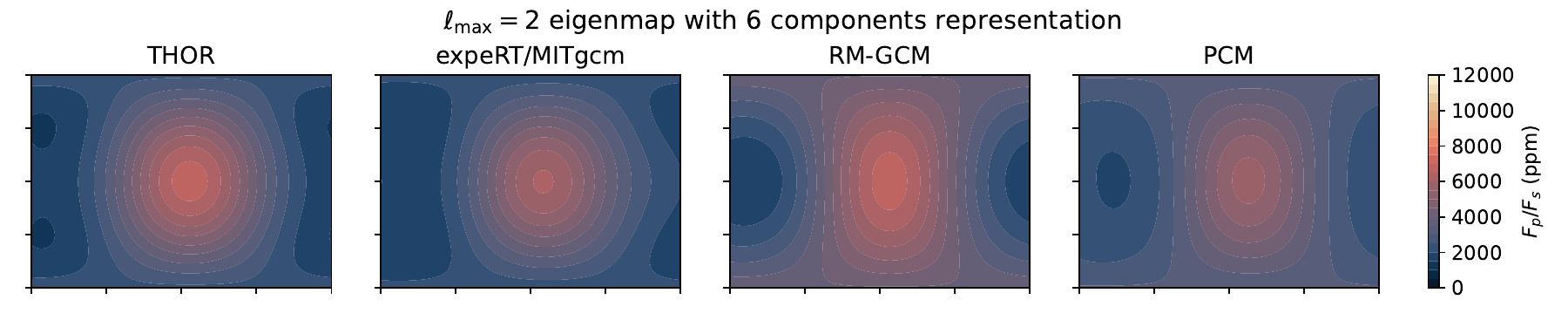}
    \includegraphics[width=\textwidth]{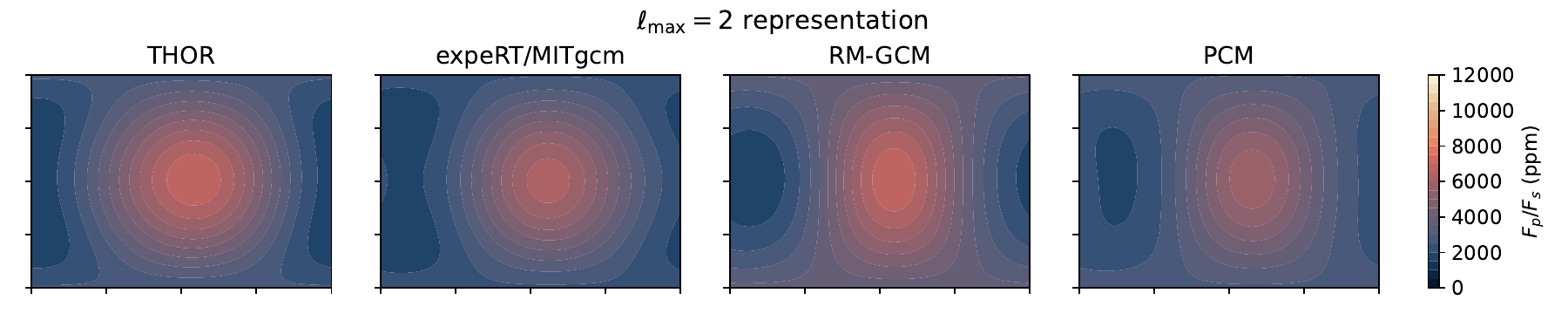}
    \caption{A comparison of four GCM simulations of WASP-43b and how their spatial distributions could appear in an eclipse map with restricted spherical harmonic order. First row: the modelled outgoing longwave radiation (OLR) in the $5$ to $10.5\,\mu$m MIRI/LRS bandpass for each of the four GCM simulations, expressed as a ratio of planetary to stellar flux. Second row: observable emission from the GCM simulations, with the null space removed as described by \citet{luger2021nullspace} and \citet{challener2023nullspace}.  Third row: the OLR represented using spherical harmonics up to $\ell_{\rm max}=3$, showing how structure is lost; this is the best map that could be achieved by a fit like the $\ell_{\rm max}=3$ fit in Figure \ref{fig:fig5_o3_map_eigenmap}. Fourth row: the OLR represented using a basis of the eigenmaps used in Figure \ref{fig:fig5_o3_map_eigenmap}. Fifth row: the OLR represented using spherical harmonics up to $\ell_{\rm max}=2$; this is the best map that could be achieved by a fit like that in Figure \ref{fig:fig3_o2_map}. All of the simulations match the dayside and nightside amplitudes of the observed phase curve and eclipse map fairly well, as well as the small eastward phase curve shift. The succession of plots here shows how using low-order spherical harmonics limits the structures that can be fitted with an eclipse map.}
    \label{fig:fig7_gcm_maps}
\end{figure*}

Figure \ref{fig:fig7_gcm_maps} shows thermal emission maps post-processed from each GCM simulation. The top row shows the thermal emission integrated from $5$ to $10.5\,\mu\text{m}$. These maps contain small-scale structure that is not present in the observed eclipse maps in Section \ref{sec:results}. A chevron-like structure is present in all the GCM maps, which is commonly attributed to the temperature structure associated with equatorial Kelvin and Rossby waves \citep{matsuno1966quasi,showman2011equatorial,lewis2022temperature}. THOR, expeRT/MITgcm, and the RM-GCM each show cold `lobes' on the nightside, which are likely associated with stationary equatorial Rossby waves \citep{matsuno1966quasi,showman2011equatorial,lewis2022temperature}. A Rossby wave-like structure also appears to dominate the dayside emission in the RM-GCM simulation. 

\begin{figure*}[!t]
    \centering
    \includegraphics[width=\textwidth]{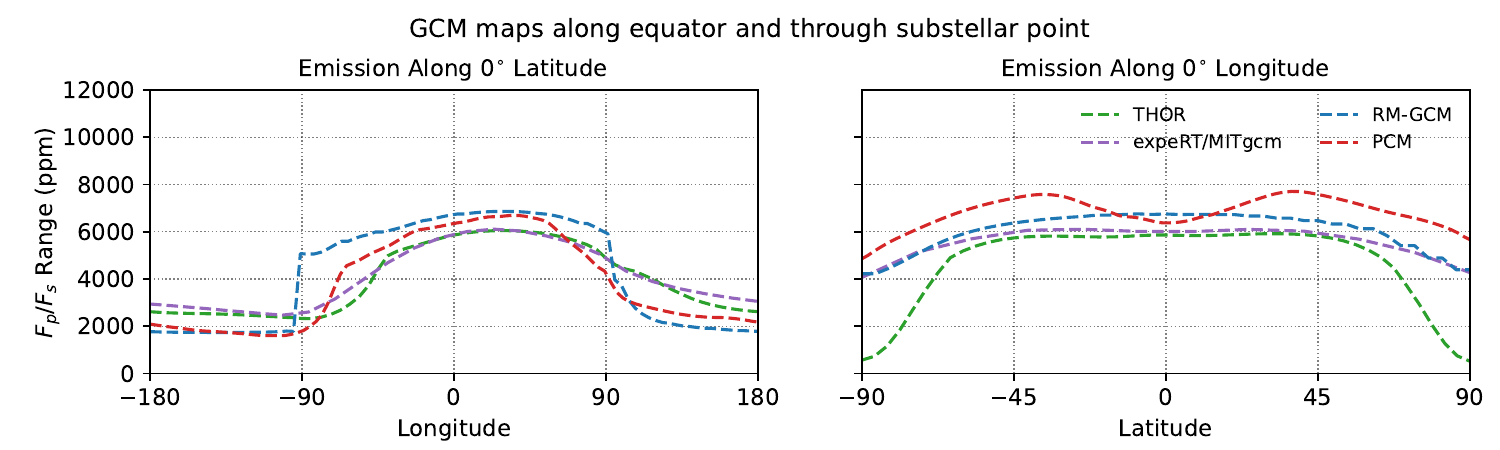}
    \includegraphics[width=\textwidth]{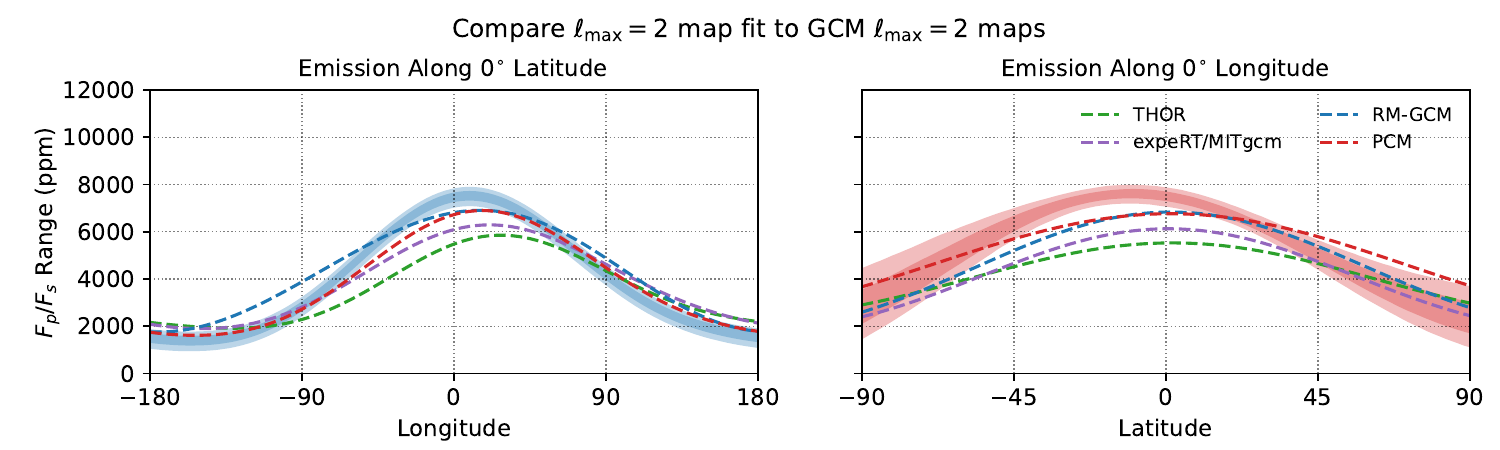}
    \includegraphics[width=\textwidth]{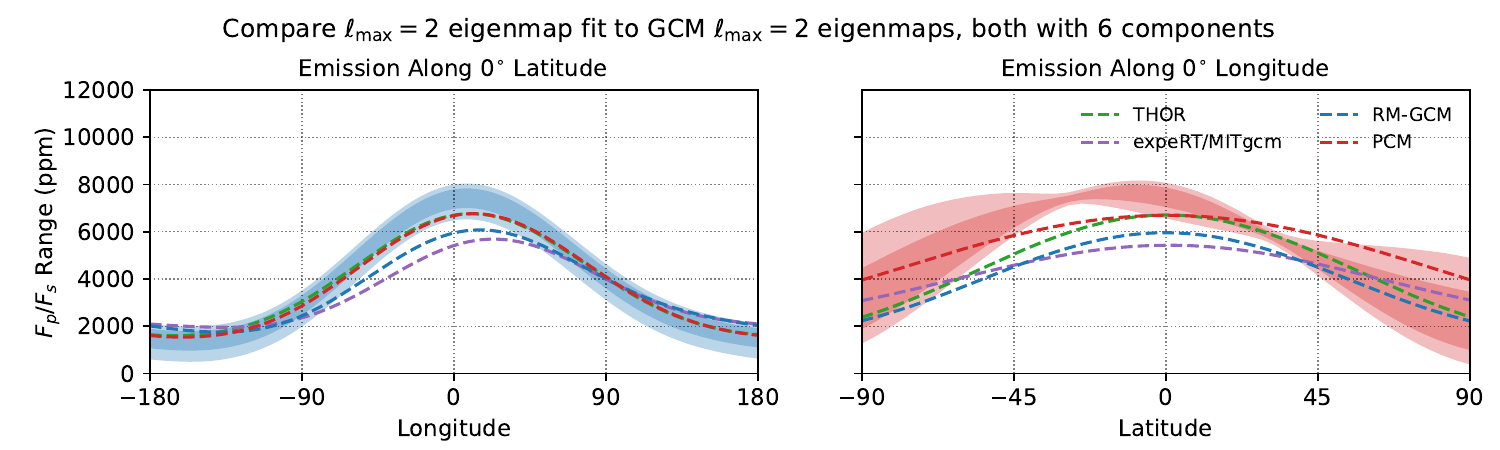}    \includegraphics[width=\textwidth]{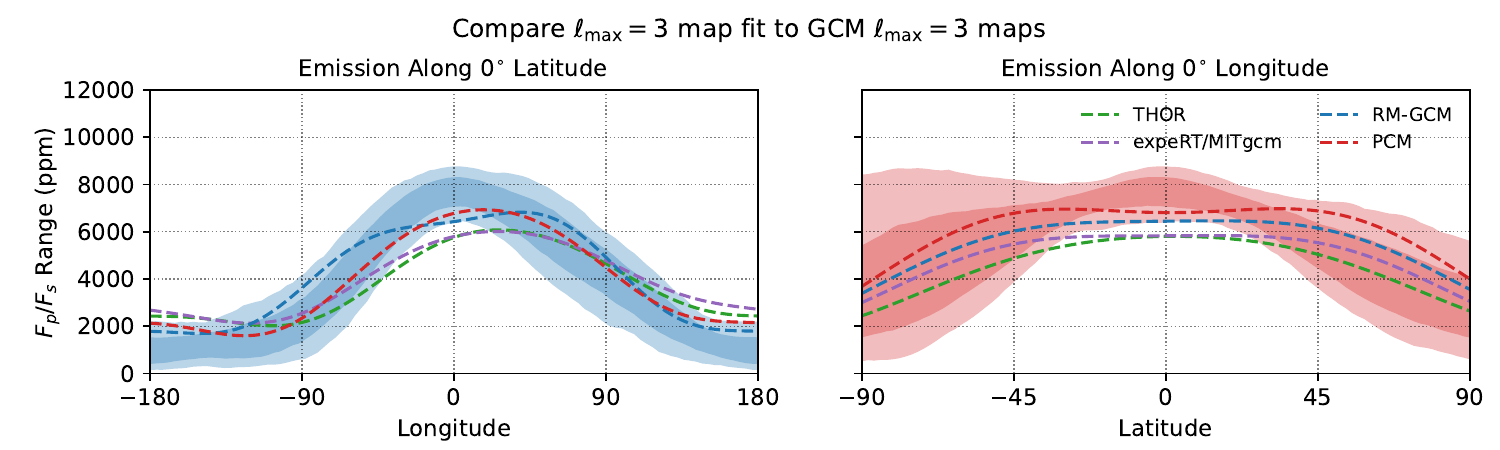}
    \caption{A comparison of the GCM simulation results in Figure \ref{fig:fig7_gcm_maps} to the eclipse maps fitted to the data in Figure \ref{fig:fig1_lightcurve}. First row: the GCM maps in the first row of Figure \ref{fig:fig7_gcm_maps}. Second row: the $\ell_{\rm max}=2$ map in Figure \ref{fig:fig3_o2_map} compared to the $\ell_{\rm max}=2$ GCM maps in the fourth row of Figure \ref{fig:fig7_gcm_maps}. Third row: the eigenmap in Figure \ref{fig:fig5_o3_map_eigenmap} compared to the eigenmap representations of the GCMs. Fourth row: the $\ell_{\rm max}=3$ map in Figure \ref{fig:fig3_o2_map} compared to the $\ell_{\rm max}=3$ GCM maps in the third row of Figure \ref{fig:fig7_gcm_maps}.}
    \label{fig:fig8_compare_gcm_obs}
\end{figure*}

\subsubsection{Observable and Null Maps}

Not all spatial brightness patterns produce a signal in phase curve space \citep{luger2021nullspace}. For instance, rotational phase curves of spatially unresolved objects (i.e., exoplanet phase curves) are insensitive to latitudinal structures. If the object is eclipsed, as is the case for WASP-43b, then the shape of eclipse ingress and egress breaks many of these degeneracies, but still leaves a ``null space'' of unobservable patterns \citep{challener2023nullspace}. This null space means that retrieved eclipse maps will not always match thermal emission output from GCM simulations. This effect may account for some of the discrepancy between the GCM maps shown in the top row of Figure \ref{fig:fig7_gcm_maps} and the observed eclipse maps in Section \ref{sec:results}. 

To illustrate this effect, we separate our GCM maps into their observable and null components following \citet{challener2023nullspace}. The second row of Figure \ref{fig:fig7_gcm_maps} shows the observable component for each GCM. The most significant difference between the observable GCM maps and the original GCM maps is the lack of spatial variation on the nightside of the observable maps. In this region, we only have information from the phase curve, so our mapping capabilities are limited to large-scale longitudinal variation. This means that the fine structure associated with, for example, nightside equatorial Rossby waves, cannot be constrained by observations. By contrast, the dayside is scanned by the eclipse, which means that some features remain observable, such as the broad shape and location of the hot-spot. While the observable maps appear to have some new structures compared to the original GCM maps, this is simply the result of removing the high-order structures of the null space. 

There are a number of differences between the observable parts of the GCMs, and the observed eclipse maps. Each observable GCM map has a large hot-spot that is offset eastwards from the substellar point, while the eclipse maps \ref{sec:results} have smaller or negligible hot-spot sifts. In addition, the dayside emission in the GCMs varies less with latitude near the equator than the emission in any of the $\ell_{\rm max}=2$ eclipse maps (Figures \ref{fig:fig3_o2_map} and \ref{fig:fig6_e1_e2_ee_map}), or the eclipse map derived using the eigenmapping method (bottom row of Figure \ref{fig:fig5_o3_map_eigenmap}).

None of the GCM maps in the top row of Figure \ref{fig:fig7_gcm_maps} show a significant latitudinal hot-spot offset, which is present in the observed $\ell_{\rm max}=2$ full phase curve map, the $\ell_{\rm max}=2$ map derived from the first eclipse only, and the eigenmap. It is unsurprising that the GCMs do not show a latitudinal offset, given that the forcing and boundary conditions for each model are hemispherically symmetric. However, the observable maps for each do have small latitudinal offsets due to asymmetries introduced by the inclined viewing angle. This implies that asymmetries could be introduced by mapping at high orders.

% Additionally, none of the observable GCM maps show a significant latitudinal hot-spot offset, which is present in the observed $\ell_{\rm max}=2$ full phase curve map, the $\ell_{\rm max}=2$ map derived from the first eclipse only, and the eigenmap. It is unsurprising that the GCMs do not show a latitudinal offset, given that the forcing and boundary conditions for each model are hemispherically symmetric. Further repeat observations would be needed to determine if the observed latitudinal offset is robust.

For the purposes of comparison to the $\ell_{\rm max}=2$ and $\ell_{\rm max}=3$ maps we fit to the observations, it is important to note that $\ell_{\rm max}=2$ and $\ell_{\rm max}=3$ maps have no null space for this observation. The non-zero null spaces identified in Figure \ref{fig:fig7_gcm_maps} are due to the high-order spherical harmonic bases used to represent the small-scale structure of the GCM results. Therefore, while the presence of a null space places a theoretical upper limit on the accuracy of eclipse mapping it should not have a direct effect on our fitted $\ell_{\rm max}=2$ and $\ell_{\rm max}=3$ maps, and should not introduce a latitudinal asymmetry at these low orders. The eigenmap basis may have some latitudinal asymmetry introduced by its calculation of the structures which contribute most strongly to the light curve.

\newpage
\subsubsection{Effect of Mapping Order}

The eclipse maps in Section \ref{sec:results} are fitted with low-order spherical harmonics up to $\ell_{\rm max}=2$ and $\ell_{\rm max}=3$, as including higher-order harmonics could lead to overfitting. This means that some realistic structures cannot be fitted accurately --- for example, a sharp brightness temperature gradient caused by the onset of cloud formation needs high-order harmonics to fit it accurately \citep{parmentier2016transitions}. To illustrate this effect, the third and fourth rows of Figure \ref{fig:fig7_gcm_maps} show emission maps from each GCM, truncated to use information from harmonics up to $\ell_{\rm max}=2$ and $\ell_{\rm max}=3$ only (computed using \texttt{starry}). 

Latitudinal and longitudinal cross-sections of the truncated maps (taken at the sub-stellar longitude, and along the equator, respectively) are shown in Figure \ref{fig:fig8_compare_gcm_obs}, where they are compared to the relevant observed eclipse maps. The eigenmaps fitted in Figure \ref{fig:fig5_o3_map_eigenmap} are also compared to each GCM expressed using a basis of the eigenmaps from this particular fit. 

Truncation of the GCM emission to $\ell_{\rm max}=2$ forces the dayside emission to be more strongly peaked on the equator than in the higher-order representations. This difference is consistent with the different latitudinal structure suggested by the $\ell_{\rm max}=2$ and $\ell_{\rm max}=3$ eclipse maps from Section \ref{sec:results} (comparing, e.g., Figures \ref{fig:fig3_o2_map} and the top row of Figure \ref{fig:fig5_o3_map_eigenmap}). The $\ell_{\rm max}=2$ fit is therefore consistent with the real WASP-43b either having i) a peaked latitudinal structure in reality, or ii) a flatter latitudinal structure in reality (like the GCMs), which is being masked by the spherical harmonic truncation. While the flatter fits in the $\ell_{\rm max}=3$ map in Figure \ref{fig:fig5_o3_map_eigenmap} are in better agreement with the GCM simulations, the $\ell=3$ modes are not constrained well enough to conclude that the flatter structure is a better representation of the ``real'' emission from WASP-43b. 

Turning to longitudinal structure, Figure \ref{fig:fig7_gcm_maps} shows that restricting the emission to $\ell_{\rm max}=2$ broadens the longitudinal structure of the hot-spot, most notably for the THOR and expeRT/MITgcm simulations. It also removes sharp gradients in emission at the terminators of the THOR and RM-GCM simulations, which are associated with the formation of clouds on the nightside. As with the latitudinal structure, these differences between $\ell_{\rm max}=2$ and $\ell_{\rm max}=3$ are consistent with those present in the eclipse maps presented in Section \ref{sec:results} and Figure \ref{fig:fig8_compare_gcm_obs}.

\begin{figure*}
    \centering
    \includegraphics[width=\textwidth]{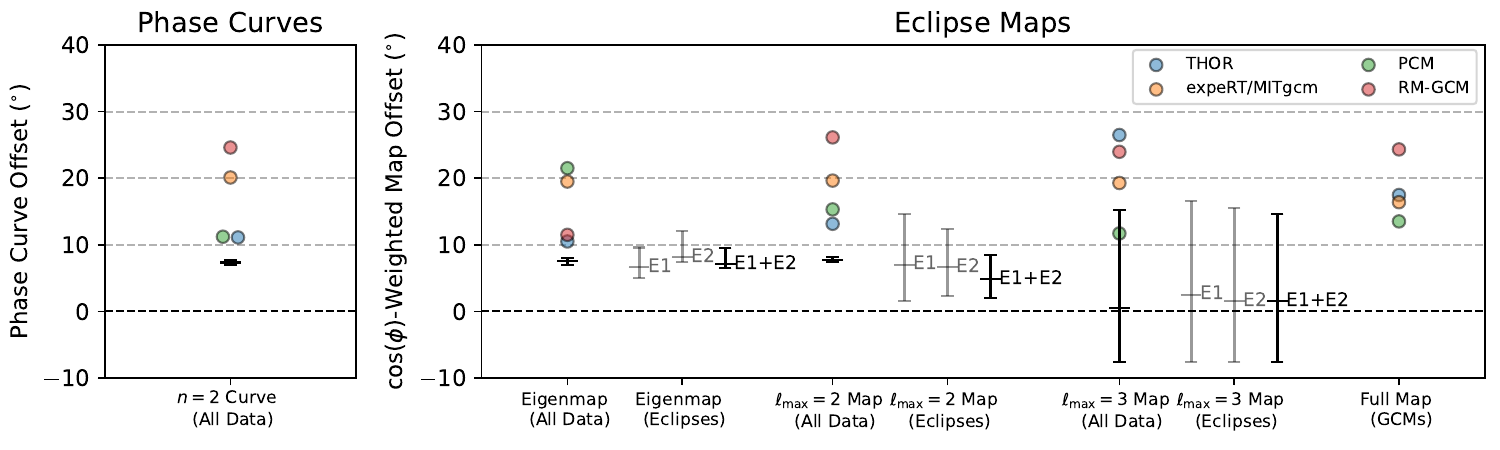}
    \caption{Left: the phase curve offset derived from the $n=2$ fit using a Fourier series model (giving the same result as \citet{bell2023wasp43b}), compared to the phase curve offset simulated from each GCM. Right: The (meridionally averaged) longitudinal offset in each 2D map. The $\ell_{\rm max}=2$ and $\ell_{\rm max}=3$ offsets (using all the dataset) are from the map fits in Figures \ref{fig:fig3_o2_map} and \ref{fig:fig5_o3_map_eigenmap}. The $\ell_{\rm max}=2$ map fits using the eclipses only correspond to the fits in Figure \ref{fig:fig6_e1_e2_ee_map} (the black point shows the fit to both eclipses, and the grey points show the fits to the first and second eclipses). The eclipse-only eigenmap fits and the eclipse-only $\ell_{\rm max}=3$ map fits are not plotted in full elsewhere. The GCM points show the $\cos\phi$-weighted offset for the GCMs truncated to $\ell_{\rm max}=2$ and $\ell_{\rm max}=3$ representations shown in Figure \ref{fig:fig7_gcm_maps}, as well as for the maps with no truncation.} 
    \label{fig:fig9_compare_offsets}
\end{figure*}

\subsection{Longitudinal Offsets}\label{sec:discussion:offsets}

The eastwards hot-spot shift of the temperature structure of tidally locked planets is one of their key observable features, traditionally derived from the phase offset of the maximum of their phase curve. This phase curve offset is not, however, identical to the actual shift of the 2D emission structure. Instead, it represents the integration of the emission over the observed hemisphere, weighted by viewing angle. Figure \ref{fig:fig9_compare_offsets} shows how the simulated phase curves for each GCM have different phase curve offsets than the offsets of their emission map, which we define as the maximum longitude $\theta_{\mathrm{max}}$ of $\int^{+\pi/2}_{-\pi/2} F_{p}(\theta,\phi) \cos\phi\ d\phi$. We refer to this averaging as ``meridional averaging'' from now on, and always consider this to be the ``longitudinal offset'' of a 2D map, for comparison to the ``peak offset'' of a phase curve.

By this measure, the THOR GCM has a phase curve offset of $11.1^{\circ}$ but an emission map offset of $17.5^{\circ}$ (meridionally averaged). Figure \ref{fig:fig9_compare_offsets} shows how this difference is especially pronounced for this simulation, due to its sharp gradients shown in the top row of Figure \ref{fig:fig7_gcm_maps}. These gradients are poorly captured by the $\ell_{\rm max}=2$ representation, so the phase curve has a very different offset, as it is dominated by low-order modes \citep{cowan2008inverting}. This also results in different emission map offsets in its $\ell_{\rm max}=2$ and $\ell_{\rm max}=3$ representations, which are $13.2^{\circ}$ and $26.5^{\circ}$ respectively. The $\ell_{\rm max}=2$ emission map offset is similar to the phase curve offset, as expected. This shows how low-order representations of emission maps can bias measurements of the large-scale offsets, as well as discard small-scale information. 

Figure \ref{fig:fig9_compare_offsets} shows that the meridionally averaged emission map offset for the $\ell_{\rm max}=2$ map in Figure \ref{fig:fig3_o2_map} is $(7.75^{+0.36}_{-0.36})^{\circ}$. This is consistent with the emission map offset for the first, second, and combined eclipses in Figure \ref{fig:fig6_e1_e2_ee_map}, which are $( 7.03 ^{+ 7.57 } _{- 5.41 })^{\circ}$, $( 6.67 ^{+ 5.77 } _{- 4.32 })^{\circ}$, and $( 4.86 ^{+ 3.6 } _{- 2.88 })^{\circ}$ respectively. The fit to the full phase curve is more precise due to the increased longitudinal information present in the out-of-eclipse phase curve, which constrains the low-order longitudinal structure very well and break degeneracies with the latitudinal structure \citep{boone2023analytic}. The $\ell_{\rm max}=3$ map fit to the whole dataset shown in Figure \ref{fig:fig5_o3_map_eigenmap} has a map offset of $( 0.50 ^{+ 14.79 } _{- 8.04 })^{\circ}$ that is consistent with the $\ell_{\rm max}=3$ eclipse-only fits, as well as with the $\ell_{\rm max}=2$ map fits. As discussed in Section \ref{sec:results:map_o3}, this map has a hot-spot offset with a median position on the substellar point, but a non-zero phase offset of around 7 degrees, consistent with the overall phase curve, showing the difference between these two metrics. The $\ell_{\rm max}=3$ fit also has much more uncertainty due to its increased degrees of mapping freedom, resulting in the large range of fitted maps in Figure \ref{fig:fig5_o3_map_eigenmap}.

The longitudinal offsets of the $n=2$ Fourier series fit, and the $\ell_{\rm max}=2$ map fit to the full dataset, appear very precise compared to the $\ell_{\rm max}=3$ map fit to the full dataset. This reflects the choice of fitting functions rather than a truly increased precision. The out-of-eclipse phase curve in Figure \ref{fig:fig1_lightcurve} is very well constrained, so the $n=2$ Fourier series model has a very precise peak as shown in Figure \ref{fig:fig9_compare_offsets}. Similarly, most of the $\ell_{\rm max}=2$ modes in the map in Figure \ref{fig:fig3_o2_map} are tightly constrained by the out-of-eclipse phase curve, resulting in an apparently more precise value in Figure \ref{fig:fig9_compare_offsets}. This precision is due to the limited fitting functions rather than actual statistical certainty. This is shown by the greater uncertainty on the $\ell_{\rm max}=3$ map in Figure \ref{fig:fig5_o3_map_eigenmap}, which has more degrees of mapping freedom. It can be consistent with the out-of-eclipse phase curve with many different maps, resulting in the larger uncertainty on its meridionally averaged longitudinal shift in Figure \ref{fig:fig9_compare_offsets}.

In summary, the derived emission map offset varies between around 0 and 20 degrees based on the mapping method used. The map offsets in the GCMs used for comparison are larger, varying between around 10 and 30 degrees depending on the mapping basis. We conclude that while the $\ell_{\rm max}=2$ map and eigenmap are the best constrained maps, they may be discarding important structures from the real map. The $\ell_{\rm max}=3$ map has access to more structures, but has too large an uncertainty to provide precise conclusions. Observations of more eclipses would provide the information needed to better constrain the $\ell=3$ modes.

\section{Conclusions}\label{sec:conclusions}

In this study, we have presented eclipse maps of the thermal emission of WASP-43b derived from a JWST MIRI/LRS phase curve. We fitted a map using $\ell_{\rm max}=2$ spherical harmonics simultaneously with the parameters of the system and the systematic parameters of the instrument. There is a clear residual mapping signal in the ingress and egress of the eclipses, which a 2D map model fits much better than a Fourier series model. This residual signal is at most 450 ppm, composed of a signal $\sim$250 ppm due to longitudinal structure and $\sim$200 ppm due to latitudinal structure. This is the first time that the magnitude of the signal of the latitudinal structure has been statistically identified in an eclipse map of an exoplanet.

Fitting this dataset with an eclipse map model derived statistically significantly different parameters to those derived with a Fourier series model as in \citet{bell2023wasp43b}. We suggest these updated parameters are more accurate due to a more accurate fit to the eclipse shapes, but note that they have little effect on the eclipse map itself, as shown in Figure \ref{fig:fig11_map_o2_fourier_params}. There were no significant degeneracies between the fitted parameters.

Figure \ref{fig:fig3_o2_map} shows that the $\ell_{\rm max}=2$ map found a small eastward hot-spot shift of $(7.75^{+0.36}_{-0.36})^{\circ}$ (defined using a meridional average), compared to an eastward phase shift of $(7.3^{+0.4}_{-0.4})^{\circ}$ derived from fitting a Fourier series to the phase curve. This $\ell_{\rm max}=2$ fit finds a sharply peaked latitudinal structure, with a small latitudinal offset of $( -10.72 ^{+ 4.14 } _{- 4.68 })^{\circ}$.

Figure \ref{fig:fig5_o3_map_eigenmap} shows an $\ell_{\rm max}=3$ eclipse map fitted with the non-mapping parameters fixed to those derived from the $\ell_{\rm max}=2$ map fit. This found a smaller but more uncertain hot-spot shift of $( 0.5 ^{+ 14.79 } _{- 8.04 })^{\circ}$ degrees east, and a flatter latitudinal structure near the equator than the $\ell_{\rm max}=2$ map. It achieved a better $\chi^{2}$ than the $\ell_{\rm max}=2$ model, but a worse BIC and AIC due to its higher number of parameters.

In addition, we fitted an ``eigenmap'' (with the orbital and systematic parameters fixed), which fitted the data with six eigenmaps with orthogonal phase curves from an $\ell_{\rm max}=2$ basis. Figure \ref{fig:fig5_o3_map_eigenmap} shows that this derived a map very similar to the original $\ell_{\rm max}=2$  eclipse map in Figure \ref{fig:fig3_o2_map}, but with fewer parameters as it discarded the mapping structures that contribute least to the phase curve. Table \ref{tab:model_comparison} shows how this model achieved the best BIC score, with a similar fit quality but fewer parameters.

As described in Section \ref{sec:results:map}, a simple estimate of brightness temperature from the observed broadband flux corresponds to a temperature at the substellar point of $( 1790.0 ^{+ 23.0 } _{- 29.0 }) $ K for the $\ell_{\rm max}=2$ map. This is consistent with the brightness temperature of the eigenmap at the substellar point of $( 1783.0 ^{+ 51.0 } _{- 70.0 }) $ K, as well as with the brightness temperature of the $\ell_{\rm max}=3$ map which is $( 1822.0 ^{+ 75.0 } _{- 92.0 }) $ K at the substellar point.

We tested the mapping information in each eclipse separately by re-fitting a map to each eclipse individually. Figure \ref{fig:fig6_e1_e2_ee_map} shows that the second eclipse produced a dayside map similar to that derived from the full phase curve, but the first eclipse produced one with a larger latitudinal offset. This may be due to increased systematic effects on the first eclipse; it could also be a real difference in thermal emission structure but we suggest that such large-scale variability over one orbit is unlikely. We recommend that those looking to use MIRI/LRS for eclipse mapping observations allow for significant settling time at the start of their observations to reduce the impact of the large instrumental systematics.

Figures \ref{fig:fig7_gcm_maps} and \ref{fig:fig8_compare_gcm_obs} compare the fitted eclipse maps to four GCM simulations from \citet{bell2023wasp43b}. In general, the GCMs match the eclipse maps well given the effect of truncating the spherical harmonic order of the maps. The $\ell_{\rm max}=2$ map is largely consistent with the $\ell_{\rm max}=2$ GCM representations. The $\ell_{\rm max}=3$ map has almost no hot-spot shift, different to the GCMs. It has a flatter latitudinal structure near the equator, which appears more consistent with the structure of the GCMs, but the higher uncertainty on the $\ell_{\rm max}=3$ map makes it difficult to compare exactly.

We conclude that there is a strong eclipse mapping signal in this MIRI/LRS observation of WASP-43b, which can strongly constrain the $\ell_{\rm max}=2$ spherical harmonic components of the planetary emission. Our fiducial map is the $\ell_{\rm max}=2$ eclipse map in Figure \ref{fig:fig3_o2_map}, fitted simultaneously with the orbital, stellar, and systematic parameters. This fiducial map finds a (meridionally averaged) hot-spot shift of $(7.75^{+0.36}_{-0.36})^{\circ}$  eastward (shown in Figure \ref{fig:fig9_compare_offsets}), compared to a shift in the peak of the phase curve of $(7.3^{+0.4}_{-0.4})^{\circ}$ eastward.

We also highlight our statistically preferred map (according to the BIC and AIC), which is the eigenmap fitted using $\ell_{\rm max}=3$ and $N_{E}=6$ basis maps; this produces a very similar map to the fiducial $\ell_{\rm max}=2$ map using fewer fitting parameters. However, the limited spatial order of the fitting functions means we may be missing important structure as shown in Figure \ref{fig:fig7_gcm_maps}. The limitation of our maps in general to $\ell_{\rm max}=2$ structures is a very model-dependent result; there may be information about smaller-scale structures in the data that the spherical harmonics are not well suited to capture. Future studies could investigate mapping methods with more degrees of freedom to relax the dependence of our conclusions on the form of the low-order spherical harmonics \citep{horne1985images}. Measuring further eclipses would allow better constraints on the smaller-scale structures. In general, we suggest that observing multiple eclipses, and ideally observing a full phase curve, is necessary to obtain reliable eclipse maps for even the best targets like WASP-43b. 

In summary, the dayside eclipse maps fitted to this dataset are generally symmetric about the equator, have a small hot-spot shift eastward from the substellar point, and vary smoothly away from the substellar point. This structure implies a weak eastward heat transport by atmospheric dynamics \citep{hammond2021rotational}, no strong latitudinal asymmetries driven by magnetic fields or atmospheric dynamics \citep{rogers2014magnetic,skinner2022modons}, and no dayside homogeneity driven by magnetic fields \citep{beltz2021exploring}. They are generally consistent with our four numerical simulations of WASP-43b, which have small eastwards hemispheric hot-spot shifts, and are symmetric about the equator. However, the eastward hot-spot shift measured by the eclipse maps is generally smaller than that predicted by the simulations. This dataset contains spectroscopic information that we averaged out in this broadband analysis; future studies could derive maps in particular wavelength bands or regions of molecular absorption, in order to measure three-dimensional temperature structure or the distribution of particular chemical species. However, the accompanying decrease in precision may weaken the mapping signal and the accuracy of the resulting maps. More precise mapping may also reveal subtler spatial features that are not present in this analysis, and mapping of planets with different properties may reveal different spatial structures on their daysides.

\begin{acknowledgements} 
This work is based on observations made with the NASA/ESA/CSA JWST. The data presented in this article were obtained from the Mikulski Archive for Space Telescopes (MAST) at the Space Telescope Science Institute, which is operated by the Association of Universities for Research in Astronomy, Inc., under NASA contract NAS 5-03127. The specific observations analyzed are associated with program JWST-ERS-01366 and can be accessed via \dataset[DOI: 10.17909/kj9a-8d81]{https://doi.org/10.17909/kj9a-8d81} Support for this program was provided by NASA through a grant from the Space Telescope Science Institute. The results reported herein benefited during the design phase from collaborations and/or information exchange within NASA’s Nexus for Exoplanet System Science (NExSS) research coordination network sponsored by NASA’s Science Mission Directorate.

M.H. acknowledges support from Christ Church, University of Oxford.
L.T. acknowledges access to the HPC resources of MesoPSL financed by the Region Ile de France and the project Equip@Meso (reference ANR-10-EQPX-29-01) of the programme Investissements d’Avenir supervised by the Agence Nationale pour la Recherche.
T.J.B.~acknowledges funding support from the NASA Next Generation Space Telescope Flight Investigations program (now JWST) via WBS 411672.07.05.05.03.02. 
\end{acknowledgements} 

\appendix 

\begin{figure*}
    \centering
    \includegraphics[width=\textwidth]{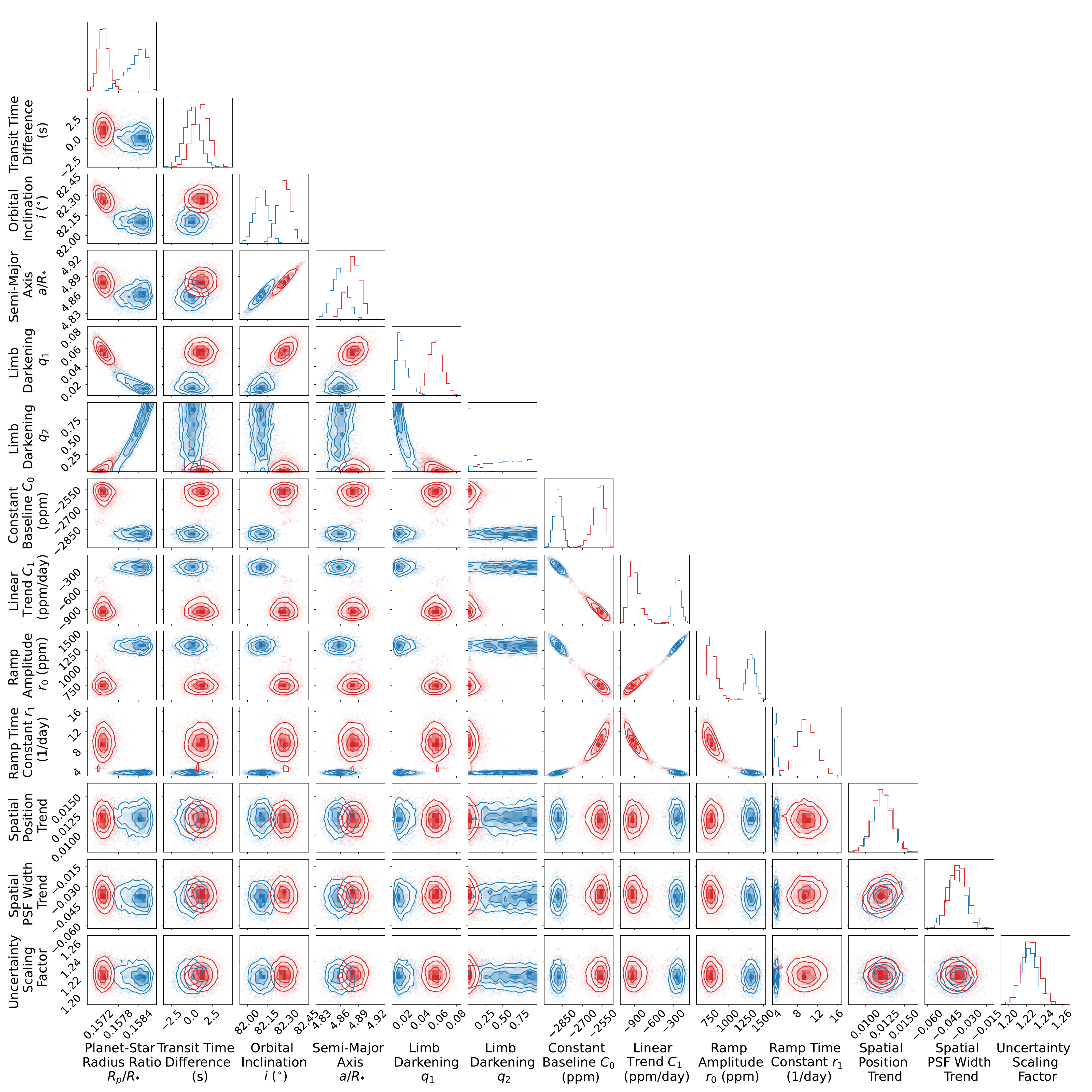}
    \caption{A corner plot of the posterior distributions of the orbital, planetary, stellar, and systematic parameters fitted to the observations in Figure \ref{fig:fig1_lightcurve}, produced with \citet{foreman2016corner}. The red posteriors correspond to the $n=2$ Fourier series model described in Section \ref{sec:results}, and the blue posteriors correspond to the $\ell_{\rm max}=2$ eclipse map model in Figure \ref{fig:fig3_o2_map}. Table \ref{tab:params_table} lists the numerical values. Some of the parameters have statistically significant differences, although Figure \ref{fig:fig11_map_o2_fourier_params} shows that these differences do not produce large changes in the eclipse map. We expect that the blue posteriors derived by the $\ell_{\rm max}=2$ eclipse map model are more accurate, due to its more accurate model of the eclipse shape.\vspace*{2cm}}
    \label{fig:fig10_corner}
\end{figure*}

\section{GCMs and post-processing}\label{ap:gcms}

We compare four GCM simulations from \citet{bell2023wasp43b} to our derived eclipse maps. The first is THOR \citep{mendoncca2016thor} (Simulation 31 in \citealp{bell2023wasp43b}), with the same model configuration as was used to simulate WASP-43b previously \citep{mendoncca2018revisiting,mendoncca2018three}. The simulation presented here uses semi-grey radiative transfer, and a simple parameterized cloud scheme on the nightside of the planet \citep{mendoncca2018revisiting}. It was run for roughly 9400 orbits, and the output data was averaged over the last 500 days. The emission from each column was calculated by post-processing the results with a multiwavelength radiative transfer model \citep{mendoncca2015new}, assuming $1{\times}$ solar metallicity and equilibrium chemical species concentration calculated with the FastChem model \citep{stock2018fastchem}.

The second is expeRT/MITgcm (Simulation 20 in \citealp{bell2023wasp43b}). This model uses the dynamical core of the MITgcm \citep{adcroft2004implementation} on a C32 cubed-sphere grid, coupled to a non-grey radiative transfer scheme based on petitRADTRANS \citep{molliere2019petitradtrans}. It follows a setup used in \citet{carone2020equatorial} and \citet{schneider2022exploring} to investigate the deep dynamics of hot Jupiters, with radiative transfer as described in \citet{schneider2022exploring}. The only difference between the configuration used here and that in \citet{schneider2022exploring} is the omission of TiO and VO. Eleven frequency bins are used, with 16 k-coefficients for each. The simulation shown here has 47 vertical levels and  $10{\times}$ solar metallicity, and was run for 1500 days, with the results shown here averaged over the last 100 days. The spectrally resolved emission was post-processed using petitRADTRANS \citep{molliere2019petitradtrans} and prt\_phasecurve \citep{schneider2022exploring} using a spectral resolution of $R=100$.

The third is the Generic Planetary Climate Model (\texttt{Generic PCM}) (Simulation 7 in \citealp{bell2023wasp43b}), which has been used to model exoplanets \citep{charnay_3d_2015,turbet_habitability_2016,teinturier2023} and the planets of the Solar System \citep{Turbet2021,Spiga2020}. The simulation presented here uses a horizontal resolution of 64$\times$48 with 40 vertical levels. The \texttt{Generic PCM} treats clouds as radiatively active tracers of fixed radii. The correlated-k radiative transfer uses 27 shortwave and 26 longwave bins, each with 16 k-coefficients. The model is run for 2000 orbits, and the data presented is an average of the model output over the last 100 days. The simulation presented here has $1{\times}$ solar metallicity and models Mg$_{2}$SiO$_{4}$ clouds with radius $1\mu$m.  The simulation was post-processed with the Pytmosph3R code \citep{falco2022toward} to calculate the spectrally resolved emission from each column, which was then integrated from 5 to 10.5 $\mu$m.

The fourth is the RM-GCM (Simulation 26 in \citealp{bell2023wasp43b}), which was adapted from the GCM of \citet{hoskins1975multi} by \citet{menou2009atmospheric}, \citet{rauscher2010three}, and \citet{rauscher2012general}, and has been applied to investigations of exoplanets using semi-grey radiative transfer and aerosol scattering \citep{roman2017modeling,roman2021clouds}. The simulation presented here has 1x solar metallicity and condensate clouds represented by aerosols as described in \citet{bell2023wasp43b}, with 50 vertical levels, and was run for over 3500 orbits. It was post-processed following \citet{zhang2017constraining} and \citet{malsky2021modeling} to produce the spectrally resolved emission.

\begin{figure*}
    \centering
    \includegraphics[width=\textwidth]{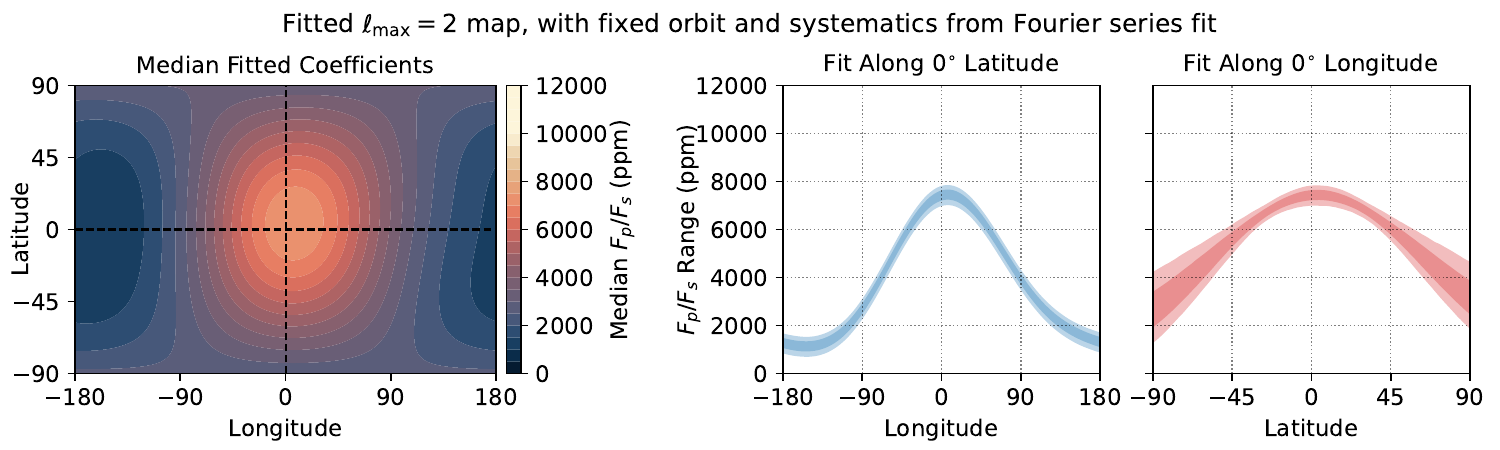}
    \caption{An $\ell_{\rm max}=2$ eclipse map fitted using the orbital and systematic parameters derived by fitting the Fourier series model, shown in Figure \ref{fig:fig10_corner} and listed in Table \ref{tab:params_table}. The map is almost identical to that in Figure \ref{fig:fig3_o2_map}, showing that the small (but statistically significant) differences between the two fits in Figure \ref{fig:fig10_corner} do not meaningfully affect the resulting $\ell_{\rm max}=2$ eclipse map.}
    \label{fig:fig11_map_o2_fourier_params}
\end{figure*}

\section{Orbital and Systematic Parameters}\label{ap:params}

Figure \ref{fig:fig10_corner} shows the orbital, planetary, stellar, and systematic parameters fitted with the $n=2$ Fourier series model described in Section \ref{sec:results}, compared to the parameters fitted with the $\ell_{\rm max}=2$ eclipse map model shown in Figure \ref{fig:fig3_o2_map}. We find no significant degeneracies between these parameters, or between these parameters and the Fourier series coefficient or map pixels used to fit the light curve shape as described in Section \ref{sec:methods} (not shown). There are some degeneracies between the parameters of the systematic model, due to inherent degeneracies between a linear slope and a long-timescale exponential ramp.

As described in Section \ref{sec:results}, the eclipse map model produces a better fit to the data (the smaller $\chi^{2}$ value listed in Table \ref{tab:model_comparison}). This is as expected, as it should provide a more realistic model of the eclipse shape. We therefore assume that the parameters derived by the eclipse map model (coloured blue in Figure \ref{fig:fig10_corner}) are more accurate. The eclipse timing is very well constrained despite the possibility for degeneracy between eclipse timing and eclipse mapping \citep{williams2006resolving}. We suggest this is due to the high precision and cadence of the observations, the independent measurement of longitudinal structure from the phase curve, and the presence of a transit in the dataset.

The Fourier series model finds a statistically significantly smaller planetary radius, higher orbital inclination, and larger semi-major axis, than the eclipse map model. The Fourier series model also finds different stellar limb darkening parameters to the eclipse mapping model, but the resulting limb darkening profile posteriors (not shown) are almost the same. This reflects degeneracies in the limb darkening model, and the limited limb darkening information available at these longer wavelengths \citep{morello2017,morello2018}. The two fits in Figure \ref{fig:fig10_corner} also find different systematic parameters, but these result in almost exactly the same systematic model. The Fourier series model fits a weak exponential ramp with a long timescale, plus a strong decreasing linear trend. The eclipse map model fits a strong exponential ramp with a short timescale, plus a weak decreasing linear trend. These two systematic models are almost identical within observational uncertainty when the linear and exponential trends are combined.

These statistically significant differences in the parameters fitted by the two models do not produce significant differences in the resulting eclipse map. Figure \ref{fig:fig11_map_o2_fourier_params} shows an $\ell_{\rm max}=2$ eclipse map fitted using orbital, planetary, stellar, and systematic parameters fixed to those derived using the $n=2$ Fourier series model (listed in Table \ref{tab:params_table}). The resulting map is very similar to that fitted simultaneously with the additional parameters in Figure \ref{fig:fig3_o2_map}, showing that the differences in parameters in Figure \ref{fig:fig10_corner} do not have a significant effect on the eclipse map. We suggest that it is still best to fit the orbital and systematic parameters simultaneously with an eclipse map model in general, to derive more accurate parameters and to search for degeneracies with the fitted map.

\bibliography{main}{}

\begin{thebibliography}{}
\expandafter\ifx\csname natexlab\endcsname\relax\def\natexlab#1{#1}\fi
\providecommand{\url}[1]{\href{#1}{#1}}
\providecommand{\dodoi}[1]{doi:~\href{http://doi.org/#1}{\nolinkurl{#1}}}
\providecommand{\doeprint}[1]{\href{http://ascl.net/#1}{\nolinkurl{http://ascl.net/#1}}}
\providecommand{\doarXiv}[1]{\href{https://arxiv.org/abs/#1}{\nolinkurl{https://arxiv.org/abs/#1}}}

\bibitem[{Adams \& Laughlin(2006)}]{adams2006long}
Adams, F.~C., \& Laughlin, G. 2006, The Astrophysical Journal, 649, 1004

\bibitem[{{Adcroft} {et~al.}(2004){Adcroft}, {Campin}, {Hill}, \& {Marshall}}]{adcroft2004implementation}
{Adcroft}, A., {Campin}, J.-M., {Hill}, C., \& {Marshall}, J. 2004, Monthly Weather Review, 132, 2845, \dodoi{10.1175/MWR2823.1}

\bibitem[{Akaike(1981)}]{akaike1981likelihood}
Akaike, H. 1981, Journal of econometrics, 16, 3

\bibitem[{{Allard} \& {Hauschildt}(1995)}]{allard1995}
{Allard}, F., \& {Hauschildt}, P.~H. 1995, \apj, 445, 433, \dodoi{10.1086/175708}

\bibitem[{{Bell} {et~al.}(2022){Bell}, {Ahrer}, {Brande}, {Carter}, {Feinstein}, {Caloca}, {Mansfield}, {Zieba}, {Piaulet}, {Benneke}, {Filippazzo}, {May}, {Roy}, {Kreidberg}, \& {Stevenson}}]{bell2022eureka}
{Bell}, T., {Ahrer}, E.-M., {Brande}, J., {et~al.} 2022, The Journal of Open Source Software, 7, 4503, \dodoi{10.21105/joss.04503}

\bibitem[{Bell {et~al.}(2024)Bell, Crouzet, Cubillos, Kreidberg, Piette, Roman, Barstow, Blecic, Carone, Coulombe, {et~al.}}]{bell2023wasp43b}
Bell, T.~J., Crouzet, N., Cubillos, P.~E., {et~al.} 2024, arXiv preprint arXiv:2401.13027

\bibitem[{Beltz {et~al.}(2021)Beltz, Rauscher, Roman, \& Guilliat}]{beltz2021exploring}
Beltz, H., Rauscher, E., Roman, M.~T., \& Guilliat, A. 2021, The Astronomical Journal, 163, 35

\bibitem[{Boone {et~al.}(2023)Boone, Grant, \& Hammond}]{boone2023analytic}
Boone, S., Grant, D., \& Hammond, M. 2023, arXiv preprint arXiv:2310.14245

\bibitem[{Bouwman {et~al.}(2023)Bouwman, Kendrew, Greene, Bell, Lagage, Schreiber, Dicken, Sloan, Espinoza, Scheithauer, {et~al.}}]{bouwman2023spectroscopic}
Bouwman, J., Kendrew, S., Greene, T.~P., {et~al.} 2023, Publications of the Astronomical Society of the Pacific, 135, 038002

\bibitem[{Burnham \& Anderson(2004)}]{burnham2004multimodel}
Burnham, K.~P., \& Anderson, D.~R. 2004, Sociological Methods \& Research, 33, 261, \dodoi{10.1177/0049124104268644}

\bibitem[{{Carone} {et~al.}(2020){Carone}, {Baeyens}, {Molli{\`e}re}, {Barth}, {Vazan}, {Decin}, {Sarkis}, {Venot}, \& {Henning}}]{carone2020equatorial}
{Carone}, L., {Baeyens}, R., {Molli{\`e}re}, P., {et~al.} 2020, \mnras, 496, 3582, \dodoi{10.1093/mnras/staa1733}

\bibitem[{{Challener} \& {Rauscher}(2022)}]{challener2022theresa}
{Challener}, R.~C., \& {Rauscher}, E. 2022, \aj, 163, 117, \dodoi{10.3847/1538-3881/ac4885}

\bibitem[{{Challener} \& {Rauscher}(2023)}]{challener2023nullspace}
---. 2023, \aj, 166, 176, \dodoi{10.3847/1538-3881/acf862}

\bibitem[{{Challener} {et~al.}(2023){Challener}, {Welbanks}, \& {McGill}}]{challener2023bringing}
{Challener}, R.~C., {Welbanks}, L., \& {McGill}, P. 2023, arXiv e-prints, arXiv:2310.03733, \dodoi{10.48550/arXiv.2310.03733}

\bibitem[{{Charnay} {et~al.}(2015){Charnay}, {Meadows}, {Misra}, {Leconte}, \& {Arney}}]{charnay_3d_2015}
{Charnay}, B., {Meadows}, V., {Misra}, A., {Leconte}, J., \& {Arney}, G. 2015, \apjl, 813, L1, \dodoi{10.1088/2041-8205/813/1/L1}

\bibitem[{{Coulombe} {et~al.}(2023){Coulombe}, {Benneke}, {Challener}, {Piette}, {Wiser}, {Mansfield}, {MacDonald}, {Beltz}, {Feinstein}, {Radica}, {Savel}, {Dos Santos}, {Bean}, {Parmentier}, {Wong}, {Rauscher}, {Komacek}, {Kempton}, {Tan}, {Hammond}, {Lewis}, {Line}, {Lee}, {Shivkumar}, {Crossfield}, {Nixon}, {Rackham}, {Wakeford}, {Welbanks}, {Zhang}, {Batalha}, {Berta-Thompson}, {Changeat}, {D{\'e}sert}, {Espinoza}, {Goyal}, {Harrington}, {Knutson}, {Kreidberg}, {L{\'o}pez-Morales}, {Shporer}, {Sing}, {Stevenson}, {Aggarwal}, {Ahrer}, {Alam}, {Bell}, {Blecic}, {Caceres}, {Carter}, {Casewell}, {Crouzet}, {Cubillos}, {Decin}, {Fortney}, {Gibson}, {Heng}, {Henning}, {Iro}, {Kendrew}, {Lagage}, {Leconte}, {Lendl}, {Lothringer}, {Mancini}, {Mikal-Evans}, {Molaverdikhani}, {Nikolov}, {Ohno}, {Palle}, {Piaulet}, {Redfield}, {Roy}, {Tsai}, {Venot}, \& {Wheatley}}]{coulombe2023wasp18b}
{Coulombe}, L.-P., {Benneke}, B., {Challener}, R., {et~al.} 2023, \nat, 620, 292, \dodoi{10.1038/s41586-023-06230-1}

\bibitem[{{Cowan} \& {Agol}(2008)}]{cowan2008inverting}
{Cowan}, N.~B., \& {Agol}, E. 2008, \apjl, 678, L129, \dodoi{10.1086/588553}

\bibitem[{{Cubillos} {et~al.}(2017){Cubillos}, {Harrington}, {Loredo}, {Lust}, {Blecic}, \& {Stemm}}]{cubillos2016correlated}
{Cubillos}, P., {Harrington}, J., {Loredo}, T.~J., {et~al.} 2017, \aj, 153, 3, \dodoi{10.3847/1538-3881/153/1/3}

\bibitem[{{de Wit} {et~al.}(2012){de Wit}, {Gillon}, {Demory}, \& {Seager}}]{de2012towards}
{de Wit}, J., {Gillon}, M., {Demory}, B.~O., \& {Seager}, S. 2012, \aap, 548, A128, \dodoi{10.1051/0004-6361/201219060}

\bibitem[{{Falco} {et~al.}(2022){Falco}, {Zingales}, {Pluriel}, \& {Leconte}}]{falco2022toward}
{Falco}, A., {Zingales}, T., {Pluriel}, W., \& {Leconte}, J. 2022, \aap, 658, A41, \dodoi{10.1051/0004-6361/202141940}

\bibitem[{{Foreman-Mackey}(2016)}]{foreman2016corner}
{Foreman-Mackey}, D. 2016, The Journal of Open Source Software, 1, 24, \dodoi{10.21105/joss.00024}

\bibitem[{Gillon {et~al.}(2012)Gillon, Triaud, Fortney, Demory, Jehin, Lendl, Magain, Kabath, Queloz, Alonso, {et~al.}}]{gillon2012trappist}
Gillon, M., Triaud, A., Fortney, J., {et~al.} 2012, Astronomy \& Astrophysics, 542, A4

\bibitem[{Gorski {et~al.}(2005)Gorski, Hivon, Banday, Wandelt, Hansen, Reinecke, \& Bartelmann}]{gorski2005healpix}
Gorski, K.~M., Hivon, E., Banday, A.~J., {et~al.} 2005, The Astrophysical Journal, 622, 759

\bibitem[{Hammond \& Lewis(2021)}]{hammond2021rotational}
Hammond, M., \& Lewis, N.~T. 2021, Proceedings of the National Academy of Sciences, 118, e2022705118

\bibitem[{{Hauschildt} {et~al.}(1999){Hauschildt}, {Allard}, \& {Baron}}]{hauschildt1999}
{Hauschildt}, P.~H., {Allard}, F., \& {Baron}, E. 1999, \apj, 512, 377, \dodoi{10.1086/306745}

\bibitem[{Hellier {et~al.}(2011)Hellier, Anderson, Cameron, Gillon, Jehin, Lendl, Maxted, Pepe, Pollacco, Queloz, {et~al.}}]{hellier2011wasp}
Hellier, C., Anderson, D., Cameron, A.~C., {et~al.} 2011, arXiv preprint arXiv:1104.2823

\bibitem[{Horne(1985)}]{horne1985images}
Horne, K. 1985, Monthly Notices of the Royal Astronomical Society, 213, 129

\bibitem[{{Hoskins} \& {Simmons}(1975)}]{hoskins1975multi}
{Hoskins}, B.~J., \& {Simmons}, A.~J. 1975, Quarterly Journal of the Royal Meteorological Society, 101, 637, \dodoi{10.1002/qj.49710142918}

\bibitem[{{Husser} {et~al.}(2013){Husser}, {Wende-von Berg}, {Dreizler}, {Homeier}, {Reiners}, {Barman}, \& {Hauschildt}}]{husser2013}
{Husser}, T.~O., {Wende-von Berg}, S., {Dreizler}, S., {et~al.} 2013, \aap, 553, A6, \dodoi{10.1051/0004-6361/201219058}

\bibitem[{Kendrew {et~al.}(2015)Kendrew, Scheithauer, Bouchet, Amiaux, Azzollini, Bouwman, Chen, Dubreuil, Fischer, Glasse, {et~al.}}]{kendrew2015mid}
Kendrew, S., Scheithauer, S., Bouchet, P., {et~al.} 2015, Publications of the Astronomical Society of the Pacific, 127, 623

\bibitem[{{Kipping}(2013)}]{kipping2013efficient}
{Kipping}, D.~M. 2013, \mnras, 435, 2152, \dodoi{10.1093/mnras/stt1435}

\bibitem[{Kokori {et~al.}(2023)Kokori, Tsiaras, Edwards, Jones, Pantelidou, Tinetti, Bewersdorff, Iliadou, Jongen, Lekkas, {et~al.}}]{kokori2023exoclock}
Kokori, A., Tsiaras, A., Edwards, B., {et~al.} 2023, The Astrophysical Journal Supplement Series, 265, 4

\bibitem[{{Lewis} \& {Hammond}(2022)}]{lewis2022temperature}
{Lewis}, N.~T., \& {Hammond}, M. 2022, \apj, 941, 171, \dodoi{10.3847/1538-4357/ac8fed}

\bibitem[{{Luger} {et~al.}(2019){Luger}, {Agol}, {Foreman-Mackey}, {Fleming}, {Lustig-Yaeger}, \& {Deitrick}}]{luger2019starry}
{Luger}, R., {Agol}, E., {Foreman-Mackey}, D., {et~al.} 2019, \aj, 157, 64, \dodoi{10.3847/1538-3881/aae8e5}

\bibitem[{{Luger} {et~al.}(2021){Luger}, {Foreman-Mackey}, {Hedges}, \& {Hogg}}]{luger2021nullspace}
{Luger}, R., {Foreman-Mackey}, D., {Hedges}, C., \& {Hogg}, D.~W. 2021, \aj, 162, 123, \dodoi{10.3847/1538-3881/abfdb8}

\bibitem[{{Majeau} {et~al.}(2012){Majeau}, {Agol}, \& {Cowan}}]{majeau2012two}
{Majeau}, C., {Agol}, E., \& {Cowan}, N.~B. 2012, \apjl, 747, L20, \dodoi{10.1088/2041-8205/747/2/L20}

\bibitem[{{Malsky} {et~al.}(2021){Malsky}, {Rauscher}, {Kempton}, {Roman}, {Long}, \& {Harada}}]{malsky2021modeling}
{Malsky}, I., {Rauscher}, E., {Kempton}, E. M.~R., {et~al.} 2021, \apj, 923, 62, \dodoi{10.3847/1538-4357/ac2a2a}

\bibitem[{{Mansfield} {et~al.}(2020){Mansfield}, {Schlawin}, {Lustig-Yaeger}, {Adams}, {Rauscher}, {Arcangeli}, {Feng}, {Gupta}, {Keating}, {Stevenson}, \& {Beatty}}]{mansfield2020eigenspectra}
{Mansfield}, M., {Schlawin}, E., {Lustig-Yaeger}, J., {et~al.} 2020, \mnras, 499, 5151, \dodoi{10.1093/mnras/staa3179}

\bibitem[{{Matsuno}(1966)}]{matsuno1966quasi}
{Matsuno}, T. 1966, Journal of the Meteorological Society of Japan, 44, 25, \dodoi{10.2151/jmsj1965.44.1_25}

\bibitem[{{McEwen} \& {Wiaux}(2011)}]{mcewen2011novel}
{McEwen}, J.~D., \& {Wiaux}, Y. 2011, IEEE Transactions on Signal Processing, 59, 5876, \dodoi{10.1109/TSP.2011.2166394}

\bibitem[{{Mendon{\c{c}}a} {et~al.}(2016){Mendon{\c{c}}a}, {Grimm}, {Grosheintz}, \& {Heng}}]{mendoncca2016thor}
{Mendon{\c{c}}a}, J.~M., {Grimm}, S.~L., {Grosheintz}, L., \& {Heng}, K. 2016, \apj, 829, 115, \dodoi{10.3847/0004-637X/829/2/115}

\bibitem[{{Mendon{\c{c}}a} {et~al.}(2018{\natexlab{a}}){Mendon{\c{c}}a}, {Malik}, {Demory}, \& {Heng}}]{mendoncca2018revisiting}
{Mendon{\c{c}}a}, J.~M., {Malik}, M., {Demory}, B.-O., \& {Heng}, K. 2018{\natexlab{a}}, \aj, 155, 150, \dodoi{10.3847/1538-3881/aaaebc}

\bibitem[{{Mendon{\c{c}}a} {et~al.}(2015){Mendon{\c{c}}a}, {Read}, {Wilson}, \& {Lee}}]{mendoncca2015new}
{Mendon{\c{c}}a}, J.~M., {Read}, P.~L., {Wilson}, C.~F., \& {Lee}, C. 2015, \planss, 105, 80, \dodoi{10.1016/j.pss.2014.11.008}

\bibitem[{{Mendon{\c{c}}a} {et~al.}(2018{\natexlab{b}}){Mendon{\c{c}}a}, {Tsai}, {Malik}, {Grimm}, \& {Heng}}]{mendoncca2018three}
{Mendon{\c{c}}a}, J.~M., {Tsai}, S.-m., {Malik}, M., {Grimm}, S.~L., \& {Heng}, K. 2018{\natexlab{b}}, \apj, 869, 107, \dodoi{10.3847/1538-4357/aaed23}

\bibitem[{{Menou} \& {Rauscher}(2009)}]{menou2009atmospheric}
{Menou}, K., \& {Rauscher}, E. 2009, \apj, 700, 887, \dodoi{10.1088/0004-637X/700/1/887}

\bibitem[{{Molli{\`e}re} {et~al.}(2019){Molli{\`e}re}, {Wardenier}, {van Boekel}, {Henning}, {Molaverdikhani}, \& {Snellen}}]{molliere2019petitradtrans}
{Molli{\`e}re}, P., {Wardenier}, J.~P., {van Boekel}, R., {et~al.} 2019, \aap, 627, A67, \dodoi{10.1051/0004-6361/201935470}

\bibitem[{{Morello}(2018)}]{morello2018}
{Morello}, G. 2018, \aj, 156, 175, \dodoi{10.3847/1538-3881/aadda4}

\bibitem[{{Morello} {et~al.}(2017){Morello}, {Tsiaras}, {Howarth}, \& {Homeier}}]{morello2017}
{Morello}, G., {Tsiaras}, A., {Howarth}, I.~D., \& {Homeier}, D. 2017, \aj, 154, 111, \dodoi{10.3847/1538-3881/aa8405}

\bibitem[{{Parmentier} {et~al.}(2016){Parmentier}, {Fortney}, {Showman}, {Morley}, \& {Marley}}]{parmentier2016transitions}
{Parmentier}, V., {Fortney}, J.~J., {Showman}, A.~P., {Morley}, C., \& {Marley}, M.~S. 2016, \apj, 828, 22, \dodoi{10.3847/0004-637X/828/1/22}

\bibitem[{Placek {et~al.}(2017)Placek, Angerhausen, \& Knuth}]{placek2017analyzing}
Placek, B., Angerhausen, D., \& Knuth, K.~H. 2017, The Astronomical Journal, 154, 154

\bibitem[{{Rauscher} \& {Menou}(2010)}]{rauscher2010three}
{Rauscher}, E., \& {Menou}, K. 2010, \apj, 714, 1334, \dodoi{10.1088/0004-637X/714/2/1334}

\bibitem[{{Rauscher} \& {Menou}(2012)}]{rauscher2012general}
---. 2012, \apj, 750, 96, \dodoi{10.1088/0004-637X/750/2/96}

\bibitem[{{Rauscher} {et~al.}(2007){Rauscher}, {Menou}, {Seager}, {Deming}, {Cho}, \& {Hansen}}]{rauscher2007toward}
{Rauscher}, E., {Menou}, K., {Seager}, S., {et~al.} 2007, \apj, 664, 1199, \dodoi{10.1086/519213}

\bibitem[{{Rauscher} {et~al.}(2018){Rauscher}, {Suri}, \& {Cowan}}]{rauscher2018eigenmap}
{Rauscher}, E., {Suri}, V., \& {Cowan}, N.~B. 2018, \aj, 156, 235, \dodoi{10.3847/1538-3881/aae57f}

\bibitem[{Rogers \& Komacek(2014)}]{rogers2014magnetic}
Rogers, T.~M., \& Komacek, T.~D. 2014, The Astrophysical Journal, 794, 132

\bibitem[{{Roman} \& {Rauscher}(2017)}]{roman2017modeling}
{Roman}, M., \& {Rauscher}, E. 2017, \apj, 850, 17, \dodoi{10.3847/1538-4357/aa8ee4}

\bibitem[{{Roman} {et~al.}(2021){Roman}, {Kempton}, {Rauscher}, {Harada}, {Bean}, \& {Stevenson}}]{roman2021clouds}
{Roman}, M.~T., {Kempton}, E. M.~R., {Rauscher}, E., {et~al.} 2021, \apj, 908, 101, \dodoi{10.3847/1538-4357/abd549}

\bibitem[{Salvatier {et~al.}(2016)Salvatier, Wiecki, \& Fonnesbeck}]{salvatier2016probabilistic}
Salvatier, J., Wiecki, T.~V., \& Fonnesbeck, C. 2016, PeerJ Computer Science, 2, e55, \dodoi{10.7717/peerj-cs.55}

\bibitem[{Scandariato {et~al.}(2022)Scandariato, Singh, Kitzmann, Lendl, Brandeker, Bruno, Bekkelien, Benz, Gutermann, Maxted, {et~al.}}]{scandariato2022phase}
Scandariato, G., Singh, V., Kitzmann, D., {et~al.} 2022, Astronomy and astrophysics, 668, A17

\bibitem[{{Schneider} {et~al.}(2022){Schneider}, {Carone}, {Decin}, {J{\o}rgensen}, {Molli{\`e}re}, {Baeyens}, {Kiefer}, \& {Helling}}]{schneider2022exploring}
{Schneider}, A.~D., {Carone}, L., {Decin}, L., {et~al.} 2022, \aap, 664, A56, \dodoi{10.1051/0004-6361/202142728}

\bibitem[{{Schwarz}(1978)}]{schwarz1978estimating}
{Schwarz}, G. 1978, Annals of Statistics, 6, 461

\bibitem[{{Showman} \& {Polvani}(2011)}]{showman2011equatorial}
{Showman}, A.~P., \& {Polvani}, L.~M. 2011, \apj, 738, 71, \dodoi{10.1088/0004-637X/738/1/71}

\bibitem[{Skinner \& Cho(2022)}]{skinner2022modons}
Skinner, J., \& Cho, J.~Y. 2022, Monthly Notices of the Royal Astronomical Society, 511, 3584

\bibitem[{{Spiga} {et~al.}(2020){Spiga}, {Guerlet}, {Millour}, {Indurain}, {Meurdesoif}, {Cabanes}, {Dubos}, {Leconte}, {Boissinot}, {Lebonnois}, {Sylvestre}, \& {Fouchet}}]{Spiga2020}
{Spiga}, A., {Guerlet}, S., {Millour}, E., {et~al.} 2020, \icarus, 335, 113377, \dodoi{10.1016/j.icarus.2019.07.011}

\bibitem[{{Stern}(1992)}]{stern1992pluto}
{Stern}, S.~A. 1992, \araa, 30, 185, \dodoi{10.1146/annurev.aa.30.090192.001153}

\bibitem[{{Stock} {et~al.}(2018){Stock}, {Kitzmann}, {Patzer}, \& {Sedlmayr}}]{stock2018fastchem}
{Stock}, J.~W., {Kitzmann}, D., {Patzer}, A. B.~C., \& {Sedlmayr}, E. 2018, \mnras, 479, 865, \dodoi{10.1093/mnras/sty1531}

\bibitem[{Stone(1977)}]{stone1977asymptotic}
Stone, M. 1977, Journal of the Royal Statistical Society: Series B (Methodological), 39, 44, \dodoi{10.1111/j.2517-6161.1977.tb01603.x}

\bibitem[{Taylor {et~al.}(2020)Taylor, Parmentier, Irwin, Aigrain, Lee, \& Krissansen-Totton}]{taylor2020understanding}
Taylor, J., Parmentier, V., Irwin, P.~G., {et~al.} 2020, Monthly Notices of the Royal Astronomical Society, 493, 4342

\bibitem[{{Teinturier} {et~al.}(2023){Teinturier}, {Charnay}, {Spiga}, {Bézard}, {Leconte}, {Mechineau}, {Ducrot}, {Millour}, \& {Clément}}]{teinturier2023}
{Teinturier}, L., {Charnay}, B., {Spiga}, A., {et~al.} 2023, \aap

\bibitem[{{Turbet} {et~al.}(2021){Turbet}, {Bolmont}, {Chaverot}, {Ehrenreich}, {Leconte}, \& {Marcq}}]{Turbet2021}
{Turbet}, M., {Bolmont}, E., {Chaverot}, G., {et~al.} 2021, \nat, 598, 276, \dodoi{10.1038/s41586-021-03873-w}

\bibitem[{Turbet {et~al.}(2016)Turbet, Leconte, Selsis, Bolmont, Forget, Ribas, Raymond, \& Anglada-Escudé}]{turbet_habitability_2016}
Turbet, M., Leconte, J., Selsis, F., {et~al.} 2016, Astronomy \& Astrophysics, 596, A112, \dodoi{10.1051/0004-6361/201629577}

\bibitem[{{Warner} {et~al.}(1971){Warner}, {Robinson}, \& {Nather}}]{warner1971measurement}
{Warner}, B., {Robinson}, E.~L., \& {Nather}, R.~E. 1971, \mnras, 154, 455, \dodoi{10.1093/mnras/154.4.455}

\bibitem[{Williams {et~al.}(2006)Williams, Charbonneau, Cooper, Showman, \& Fortney}]{williams2006resolving}
Williams, P.~K., Charbonneau, D., Cooper, C.~S., Showman, A.~P., \& Fortney, J.~J. 2006, The Astrophysical Journal, 649, 1020

\bibitem[{Yang {et~al.}(2023)Yang, Irwin, \& Barstow}]{yang2023testing}
Yang, J., Irwin, P.~G., \& Barstow, J.~K. 2023, Monthly Notices of the Royal Astronomical Society, 525, 5146, \dodoi{10.1093/mnras/stad2555}

\bibitem[{{Zhang} {et~al.}(2017){Zhang}, {Kempton}, \& {Rauscher}}]{zhang2017constraining}
{Zhang}, J., {Kempton}, E. M.~R., \& {Rauscher}, E. 2017, \apj, 851, 84, \dodoi{10.3847/1538-4357/aa9891}

\end{thebibliography}
\bibliographystyle{aasjournal}

\end{document}